%% file: vis_main.tex
\documentclass{vgtc}                          % final (conference style)
\ifpdf%                                % if we use pdflatex
  \pdfoutput=1\relax                   % create PDFs from pdfLaTeX
  \usepackage{graphicx}                % allow us to embed graphics files
  \DeclareGraphicsExtensions{.pdf,.png,.jpg,.jpeg} % for pdflatex we expect .pdf, .png, or .jpg files
\else%                                 % else we use pure latex
  \usepackage{graphicx}                % allow us to embed graphics files
  \DeclareGraphicsExtensions{.eps}     % for pure latex we expect eps files
\fi%

\graphicspath{{figures/}{pictures/}{images/}{./}} % where to search for the images
\usepackage{microtype}                 % use micro-typography (slightly more compact, better to read)
\PassOptionsToPackage{warn}{textcomp}  % to address font issues with \textrightarrow
\usepackage{textcomp}                  % use better special symbols
\usepackage{mathptmx}                  % use matching math font
\usepackage{times}                     % we use Times as the main font
\usepackage{cite}                      % needed to automatically sort the references
\usepackage{tabu}                      % only used for the table example
\usepackage{booktabs}                  % only used for the table example

\usepackage[usenames, dvipsnames]{color}
\usepackage[dvipsnames]{xcolor}

\usepackage{hyperref}
\usepackage{color}
\usepackage{enumitem}
\usepackage{graphicx}
\usepackage{fixltx2e}

\usepackage{multicol}
\usepackage{amsmath}
\usepackage{amssymb}
\usepackage{amsfonts}
\usepackage[font={scriptsize,sf}]{subfig}
\usepackage{floatrow}
\usepackage{bm} % for bold for vector
\usepackage{helvet}
\usepackage{algorithm}
\usepackage[noend]{algcompatible}
\usepackage{soul}
\usepackage{tikz}
\usepackage{pgfplots}
\usepackage{siunitx} % for SI unit
\usepackage{dblfloatfix}
\usepackage{xspace}
\newcommand{\eg}{\textit{e.g.,}\xspace}
\newcommand{\ie}{\textit{i.e.,}\xspace}
\newcommand{\etc}{\textit{etc.}\xspace}
\newcommand{\etal}{\textit{et al.}\xspace}

\usepackage{enumitem}

\newcommand{\para}[1]{\vspace{0.4em}\noindent{\textbf{#1}}}

\usepackage[mathcal]{euscript}

\newcommand{\ngf}{$\mathbf{G}$}
\newcommand{\ndg}{$\mathcal{G}$}
\newcommand{\ndf}{$\mathcal{D}$}

\newcommand{\df}{\ndf\xspace}
\newcommand{\dg}{\ndg\xspace}
\newcommand{\gf}{\ngf\xspace}

\newcommand{\gfi}{\ngf$_i$\xspace}
\newcommand{\dfi}{\ndf$_i$\xspace}
\newcommand{\dgi}{\ndg$_i$\xspace}
\newcommand{\dgj}{\ndg$_j$\xspace}
\newcommand{\dgci}{\ndg$^{\text{cct}_i}$\xspace}

\newcommand{\gfce}{\ngf$^{\text{cct}_\text{E}}$\xspace}

\newcommand{\gfse}{\ngf$^{\text{sg}_\text{E}}$\xspace}
\newcommand{\dgce}{\ndg$^{\text{cct}_\text{E}}$\xspace}
\newcommand{\dgge}{\ndg$^{\text{cg}_\text{E}}$\xspace}
\newcommand{\dgse}{\ndg$^{\text{sg}_\text{E}}$\xspace}
\newcommand{\dfe}{\ndf$_\text{E}$\xspace}

\definecolor{gray}{RGB}{191,191,191}
\definecolor{orange}{RGB}{244,177,131}
\definecolor{green}{RGB}{169,209,142}
\definecolor{myblue}{RGB}{156,195,230}

\newcommand{\pcode}[1]{\texttt{#1}}

\onlineid{1212}

\vgtccategory{Research}
\vgtcpapertype{application/design study}

\vgtcinsertpkg

\title{Scalable Comparative Visualization of Ensembles of Call Graphs}

\author{
    \hspace*{30pt}Suraj P. Kesavan\thanks{e-mail: \{spkesavan, klma\}@ucdavis.edu}\\
    \scriptsize \hspace*{30pt}University of California, Davis
\and 
    \hspace*{10pt}Harsh Bhatia\thanks{e-mail: \{bhatia4, gamblin2, bremer5\}@llnl.gov}\\
    \scriptsize \hspace*{10pt}Lawrence Livermore National Laboratory
\and 
    \hspace*{-20pt}Abhinav Bhatele\thanks{e-mail:bhatele@cs.umd.edu}\\
    \scriptsize \hspace*{-20pt}University of Maryland, College Park
\and 
    \hspace*{0pt}Todd Gamblin\footnotemark[2]\\
    \scriptsize \hspace*{0pt}Lawrence Livermore National Laboratory
    \vspace{-12pt}
\and
    \hspace*{-10pt}Peer-Timo Bremer\footnotemark[2]\\
    \scriptsize \hspace*{-10pt}Lawrence Livermore National Laboratory
    \vspace{-12pt}
\and 
    \hspace*{-10pt}Kwan-Liu Ma\footnotemark[1]\\
    \scriptsize \hspace*{-10pt}University of California, Davis\hspace*{1pt}
    \vspace{-12pt}
}

\abstract{\input{0_abstract.tex}}

\CCScatlist{
  \CCScatTwelve{Human-centered computing}{Visualization}{Visualization techniques}{Dendrograms};
  \CCScatTwelve{Human-centered computing}{Visualization}{Visualization application domains}{Information visualization}
}

\begin{document}


\firstsection{Introduction}
\maketitle

\input{f1_background}
\input{1_introduction}
\input{2_background}

\input{3_relatedwork}

\input{4_motivation}
\input{5_visual_interface}
\input{f4_lulesh-compare}
\input{f5_lulesh_summary}
\input{6_casestudy}

%\input{6_cs_osu}
\input{f6_osu_load}

\input{7_discussion}

%\input{8_conclusion}

%\section{Bibliography Instructions}
\acknowledgments{
We are grateful to Olga Pierce for their feedback that improved 
the content and readability of this work. This work was performed 
under the auspices of the U.S. Department of Energy by Lawrence 
Livermore National Laboratory (LLNL) under contract DE-AC52-07NA27344. 
LLNL-CONF-809459.
}

%\clearpage
\bibliographystyle{abbrv}

\bibliography{ref}
\end{document}

%% file: 0_abstract.tex
Optimizing the performance of large-scale parallel codes is critical for
efficient utilization of computing resources. Code developers often explore
various execution parameters, such as hardware configurations, system software
choices, and application parameters, and are interested in detecting and
understanding bottlenecks in different executions.  They often collect
hierarchical performance profiles represented as call graphs, which combine
performance metrics with their execution contexts. 
%requires detecting and understanding bottlenecks in the execution of the
%codes.  % Call graphs are succinct representations to study parallel codes
%since they combine the performance metrics along with their execution
%contexts.  % To cope with ever-increasing computational needs, code developers
%often test various execution modes, such as machine environments, architecture
%choices, and MPI parameters, consequently generating ensembles of call graphs.
%The ability to compare call graphs and assess the potential consequences of
%different choices is thus crucial. 
The crucial task of exploring multiple call graphs together is tedious and
challenging because of the many structural differences in the execution
contexts and significant variability in the collected performance metrics (\eg
execution runtime). In this paper, we present an enhanced version of CallFlow
to support the exploration of ensembles of call graphs using new types of
visualizations, analysis, graph operations, and features.  We introduce
\emph{ensemble-Sankey}, a new visual design that combines the strengths of
resource-flow (Sankey) and box-plot visualization techniques.  Whereas the
resource-flow visualization can easily and intuitively describe the graphical
nature of the call graph, the box plots overlaid on the nodes of Sankey convey
the performance variability within the ensemble.
%design based on the flow-based visualization coupled with a box-plot like
%visualization to highlight the variability among nodes of interest.  Our
%interactive visual interface follows a popular visual comparison framework
%comprising of identifying appropriate visual targets, determining actions,
%formulating strategies, and designing the visual interface. 
Our interactive visual interface provides linked views to help explore
ensembles of call graphs, \eg by facilitating the analysis of structural
differences, and identifying similar or distinct call graphs.
% , and find the best execution parameters.
We demonstrate the effectiveness and usefulness of our design through case
studies on large-scale parallel codes.
\vspace{-0.7em}

%% file: f1_background.tex
\begin{figure*}[!t]
        %\captionsetup{farskip=0pt}% <--- no gap at the top
        \centering
        \subfloat[Calling Contexts]{
                \includegraphics[height=3.8cm, keepaspectratio]{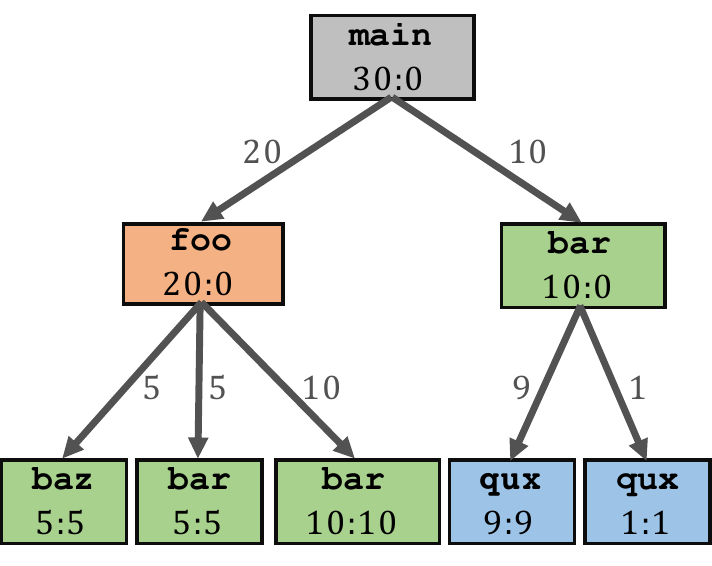}
                \label{fig:calltree}
        }
        \hfill
        \subfloat[Calling Context Tree]{
                \includegraphics[height=3.8cm, keepaspectratio]{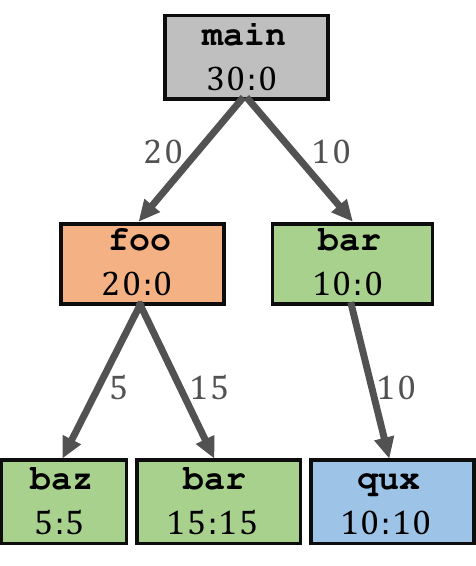}
                \label{fig:cct}
        }
        \hfill
        \subfloat[Call Graph]{
                \includegraphics[height=3.8cm, keepaspectratio]{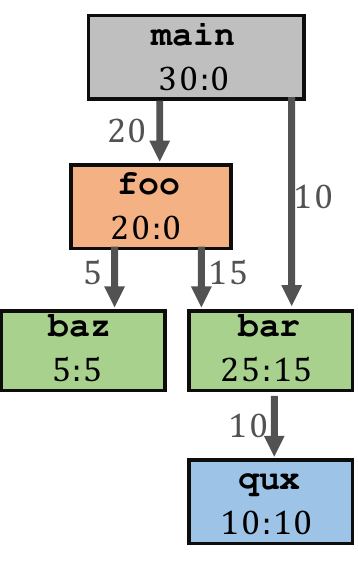}
                \label{fig:callgraph}
        }
        \hfill
        \subfloat[Super Graph]{
                \includegraphics[height=3.8cm, keepaspectratio]{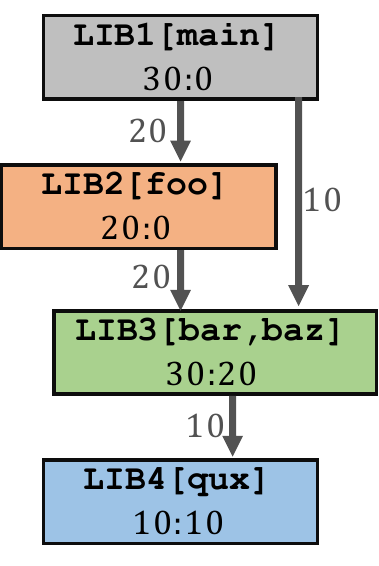}
                \label{fig:supergraph}
        }
        \vspace{-0.75em}
        \caption{Performance profiles of applications are captured as their calling
        contexts (a), which represent  call sites (functions) and their callees all the
        way up to \texttt{main}, along with performance metrics, \eg runtimes.
        CCTs (b) and call graphs (c) are simplifications of calling contexts created by
        aggregating call sites representing the same function.
        A super graph (d) introduces additional simplification through semantic
        aggregation, \eg based on the libraries they belong to.
        The nodes are labeled by the function/library name and
                {inclusive}:{exclusive} runtimes, and colors represent the nodes by library. Each edge
        is labeled with the amount of ``resource'' (exclusive runtime) flowing
        through.        \vspace{-1.5em}
\label{fig:background}}
        \end{figure*}

%% file: 1_introduction.tex
%
%% ---------------------------------------------------------------------
%% 1. Introduce the problems of comparing multiple DAGs like callgraphs.
%% 2. Simulations are used to perform experiments on the supercomputer.
%% 3. Often times, the engineers are required to improve the performance of the simulation and they use profiling the code as the method to measure different properties of a run. The profiling step introduces callgraphs.
%% 4. Current tools such as Callflow , only focus on visual analytics for a single run.
%% 5. Traditional methods for visual comparison often are not effective, like side-by-side viewing, stacked profiles in 3D.
%% 6. We introduce a visual analytic toolkit that characterizes the difference between runs and focus on representing this information to help scientists find performance bottlenecks in their simulation code.
%
%% Motivate with respect to HPC first and then tell.
%% Talk about usefulness with respect to HPC application and explain how callgraphs are used for debugging.}}
%% ---------------------------------------------------------------------

Large-scale computational resources enable scientists to explore
complex scientific phenomena and advance the frontiers of human
understanding in many areas, such as medicine~\cite{mummi2019,
smith2020repurposing} and astronomy~\cite{norman2018simulating}, 
% Nevertheless, this increased reliance on large-scale supercomputers comes
% at great cost, both economic and environmental.
%This involves studying scientific phenomena 
through computational science simulations. 
In order to maximize the amount of ``science per dollar'' and
``insights per Watt,'' computational scientists and high performance
computing (HPC) experts continuously strive towards optimizing the
performance of their simulation codes.

The key to improving the performance of such large-scale parallel codes is to
understand performance bottlenecks and identify places in the code to fix
such bottlenecks.  For example, a simulation code may utilize computing
resources inefficiently, possibly due to a suboptimal implementation,  
such as inefficient file I/O or unnecessary communication between
processing elements. Diagnosing the true causes of performance bottlenecks
typically requires a simultaneous understanding of both performance and
the place in code where the performance data was
gathered or can be attributed to --- a task typically performed by recording
and analyzing the execution of different parts of the program through
\emph{profiling}.

There exist several profiling tools~\cite{shende2006tau, knupfer2008vampir,
adhianto2010hpctoolkit, boehme2016caliper} that record the application's
\emph{calling context} (\ie the call path from each profiled function to
program main) and the associated performance metrics (\eg execution time and
memory usage).  Combining the calling contexts at different call sites forms
a single hierarchy, referred to as the \emph{calling context tree
(CCT)}~\cite{ammons1997exploiting}.  A CCT can be further aggregated based
on semantic information (\eg function names) to form \textit{call
graphs}~\cite{graham1982gprof}.  Call graphs provide a succinct and powerful
representation of program execution --- nodes of a call graph represent
functions, and the edges represent the caller-callee relationship between
functions.  Studying call graphs helps gain insights into
the execution of an application and identify potential bottlenecks and
strategies for improvement.

Several visualizations and analysis tools have been
developed~\cite{mohr2003kojak, shende2006tau, adhianto2010hpctoolkit,
geimer2010scalasca, isaacs2014state, devkota2018cfgexplorer,
nguyen2019callflow} to assist the exploration of CCTs and call graphs.
However, most of these tools do not support exploring and comparing multiple
call graphs simultaneously.  In particular, to identify optimal execution parameters and configurations,
experts often conduct a variety of test runs for
varying hardware configurations, system software choices, as
well as application parameters, resulting in large ensembles of call graphs.
%supercomputing applications aim to improve execution modes, such as machine
%environment and/or MPI parameters, as well as better utilize heterogeneous
%components like CPUs and GPUs.  % To identify optimal execution modes, HPC
%specialists often perform a variety of test runs and evaluate large
%ensembles of executions' calling contexts.
Comparing the call graphs of several executions, therefore, becomes
essential to reveal the differences in both the performance metrics and the
calling structure. However, existing call graph visualization tools either
lack the support for comparative analysis or provide only primitive
functionalities, \eg juxtaposed comparison~\cite{nguyen2019callflow}, making
an effective evaluation of hundreds of call graphs almost infeasible.
Alternative approaches, such as statistical summaries, may be used for
computing predefined metrics, but the lack of interactive visual analytic
support severely limits the experts' ability to understand new,
unanticipated causes of bottlenecks.

Nevertheless, a recently developed tool,
CallFlow~\cite{nguyen2019callflow}, opens new opportunities for supporting
scalable and interactive visual exploration of ensembles of call
graphs.  CallFlow visualizes call graphs using Sankey
diagrams~\cite{Riehmann2005InteractiveSD} to indicate the flow and
distribution of resources of interest, \eg execution runtime. CallFlow
visualization couples performance metrics with \emph{graph operations}, such
as filtering or aggregation, to visualize call graphs at user-desired
details.  CallFlow was aimed at inspecting individual call graphs, and the
comparison of multiple call graphs required the user to analyze them
individually. 

To support interactive visual exploration of large ensembles of call graphs,
we present an {enhanced} version of CallFlow. This version supports an
ensemble mode (multiple call graphs), a comparison mode (between two call
graphs or between a selected call graph and the ensemble), and
metrics to identify similar/dissimilar ensemble members.

% ----------------------------------------------------------------------------
\para{Contributions.} Our main contributions to the visualization community
include a novel visual design that enables the exploration of ensembles
of call graphs, and an interactive visual analytic tool that demonstrates our
design. In particular, our contributions are as follows.
\begin{itemize}[nosep,leftmargin=*]
\item We introduce a new visual design, \emph{ensemble-Sankey}, which combines
the strengths of resource-flow (Sankey) and box-plot visualization techniques,
facilitating the visualization of large ensembles of resource-flow graphs. 
\item Focusing on the application at hand, we present an enhanced version of
CallFlow to support exploration of ensembles of call graphs using new types of
visualizations, analysis, graph operations, and features.
%\vspace{-0.05in}
\begin{itemize}[nosep, leftmargin=*]
\item The \emph{ensemble view}, built upon ensemble-Sankey design, provides a
complete, high-level visualization of the ensemble. This view can be used to
understand the overall distribution of the concerned profiles as well as
identify the behavior of individual runs. 
\item The \emph{module hierarchy view} augments the ensemble view by
adding execution details within a selected node using
icicle plots.
\item We complement direct visualizations of call graphs with metrics for
identifying similar call graphs and interactively selecting the visualization
to show only the call graphs of user's interest.
\item We improve the performance of graph operations on large CCTs, such as
filtering, aggregation, and splitting, to support ensembles.  The new CallFlow
supports multiple data formats, including Caliper~\cite{boehme2016caliper},
HPCToolkit~\cite{adhianto2010hpctoolkit}, and gprof~\cite{graham1982gprof}.
\end{itemize}
\end{itemize}

%% file: 2_background.tex
\section{Terms and Definitions}

Sampled profiles, such as those generated by gprof~\cite{graham1982gprof},
HPCToolkit~\cite{adhianto2010hpctoolkit}, and Caliper~\cite{boehme2016caliper}
contain a variety of performance data. At every sampling point, the profiler
walks through the execution stack of the program and identifies the full
calling context at that call site. Two kinds of information are recorded:
\emph{contextual information}, \ie the current line of code, file name, the
call path, the process ID, and \emph{performance metrics} such as the number of
CPU cycles elapsed, number of floating-point operations, or branch misses
occurred since the last sample. Poor application performance or bottlenecks
typically refer to large runtimes consumed at various call sites. Two types of
timings are usually recorded: \emph{exclusive runtime}, \ie the time consumed
by a given function, and \emph{inclusive runtime}, \ie the recursively
accumulated time consumed by a given function and all its callees.

Given the raw profile data, the call paths (see~\autoref{fig:calltree})
can then be aggregated to form a \emph{calling context tree} (CCT), as shown in
\autoref{fig:cct}.  Each unique invocation (by call path) of a function becomes
a node in the CCT with the corresponding performance metrics aggregated across
all invocations, and the path from a given node to the root of the tree
represents a distinct calling context.  CCTs reduce the number of repeating
call sites with the same contextual information and are especially useful for
large applications, \eg with several hundred MPI processes that have the same
calling context.  A CCT can be further aggregated to provide information
concisely into a \emph{call graph}, which is constructed by merging the call
sites that represent the same function. For example, the call graph in
\autoref{fig:callgraph} combines all call sites of the function \texttt{bar}
from different calling contexts.  A call graph contains a set of directed
edges, each connecting a callee with its calling function. Although call
graphs provide an accurate and reduced representation of calling contexts,
their visual analysis is nevertheless severely constrained by their scale,
especially for large applications with calls to multiple libraries that may
lead to hundreds of call sites.

To simplify the visual analysis, CallFlow~\cite{nguyen2019callflow}
introduced  a \textit{super graph}, created by aggregating the
nodes of a call graph based on tool- or user-defined semantic attributes, \eg the library name,
module name, or file name. 
\autoref{fig:supergraph} shows the higher-level representation given by super graphs.
%, and comprises two kinds of nodes: \textit{super node}, which represent a
%collection call site which share the meta-attribute, and \textit{meta node},
%which  call sites.
\emph{Supernodes} (the nodes in a super graph)
represent collected call sites with a shared semantic attribute.  For example,
the supernode \texttt{lib3} in \autoref{fig:supergraph} combines the call sites
\texttt{bar} and \texttt{baz} that are part of the same library.  Although super
graphs are useful for exploring large-scale call graphs, the semantic
aggregation leads to loss of detailed information within a supernode, \eg
the calling context and the performance metrics of the functions within a given
library. In this work, we introduce new functionality to CallFlow that
preserves this information, which can be revealed to the user upon request via
the \emph{supernode hierarchy view} (see \autoref{sec:modulehierarchy}).

%% file: 3_relatedwork.tex
\section{Related Work}

To reduce visual complexity and improve readability, graph data is often
visualized as trees~\cite{ahn2009scalable, isaacs2018preserving}, including the
case of CCTs and call graphs~\cite{adamoli2010trevis, nguyen2016vipact,
bergel2017visual, nguyen2019callflow}. Here, we discuss related work 
in the area of 
call graphs, tree ensembles, and graphs in general.

%\subsection{Visualization of CCTs and Callgraphs}
% Callgraph works:
% 1. Trevis: A context tree visualization  analysis  framework  and  its  use  for  classifying performance  failure  reports. - Matrix.
% 2. Kojak - a tool set for automatic performance  analysis  of  parallel  programs. (node -encoding)
%     a. ~\cite{mohrkojak}
%     b. uses an expandable tree as the layout
% 3. A tool suite for simulation based analysis of memory access behavior.
%     a. ~\cite{weidendorfer2004tool}
%     b. Uses a tree representation.
% 4. Score-P: A unified performance measurement system for petascale applications.
% 5. HPCtoolkit
% 6. Scalesca.

\para{Visualization of CCTs and call graphs.}
Interactive visualization tools for analyzing the performance of large-scale
supercomputers have seen significant research and development efforts in recent
years~\cite{isaacs2014state}.
Focusing specifically on CCTs, one class of tools~\cite{mohr2003kojak,
shende2006tau, adhianto2010hpctoolkit, geimer2010scalasca} visualize them as
collapsible trees, where the user can toggle to show/hide functions as well as
sort by the attributes (\eg inclusive runtime). % or memory consumption).
Several call graph visualization tools employ a node-link layout technique
using a graph drawing software like GraphViz~\cite{ellson2004graphviz} and use
additional node encoding techniques to represent the data
attributes~\cite{mohr2003kojak, derose2007detecting, nguyen2016vipact}.
%
%For example, DeRose \etal~\cite{derose2007detecting} superpose a histogram
%onto the node to show imbalances between processes, and Nguyen
%\etal~\cite{nguyen2016vipact} encode runtime variation among processes to
%indicate the anomalies.
% I feel it might not be needed.  Bohnet and D\"ollner~\cite{bohnet2006visual}
% identify and embed the derived features in the data.
%
% Maybe mention something about interactivity.
However, it is well known that node-link layouts lack scalability when dealing
with large-scale graphs~\cite{Ghoniem2005}.
Additionally, the node-link layouts lack efficient user interactivity, since
the user might require to toggle functions several times for investigation,
especially for deep call stacks, and cannot directly support exploring
ensembles of CCTs.
%a call graph that possesses a deeper call stack due to calls from multiple
%libraries or modules.  do not scale easily to large trees (i.e., high number
%of functions) and are limited in their interactivity.

To address scalability and interactivity issues, HPCToolkit's
hpcviewer~\cite{adhianto2010hpctoolkit} employs additional strategies like
automatic hot path extraction within a chosen hierarchy, flattening of calling
structure, and zooming interactions to expand the relevant portion of the CCT.
Likewise, CallFlow~\cite{nguyen2019callflow} integrates different graph
splitting operations to allow the user choose the granularity of the
visualization (\ie CCT, call graph, or super graph).
CallFlow employs a Sankey diagram to encode the net inclusive time a function
or a module spends in an execution.
%
%Alternatively, Xie \etal~\cite{xie2018visual} learn the structural features
%using an anomaly behavior detection model.
%%
%The derived model embeddings are projected and their interface allows the user
%to select the interesting graphs, upon which the user can perform additional
%structural analysis.

\para{Ensembles of call graphs.} 
Supporting visual comparison of multiple call graphs that addresses the specific
needs of HPC experts would enable improved diagnosis of issues, especially with
the trend to use adaptive execution modes. Williams \etal~\cite{williams2019visualizing} amplified
the importance of comparison, mainly for execution graphs (where each edge
represents the dependency between tasks).
The authors also proposed a visualization tool that compare only two
execution graphs at any time by employing node-encodings to show the differences
in the execution graphs.
To compare several call graphs, Trevis~\cite{adamoli2010trevis}
used a matrix view to visualize pairwise similarity between graphs using a
variety of distance measures.  However, not only does the similarity matrix
suffer from scalability issues, it often requires the user to perform pairwise
comparisons~\cite{andrews2009visual} by matching individual runs.
%
%
%There also exist tools to visualize similarity of call graphs.
%%
%For example, Trevis~\cite{adamoli2010trevis} employs a matrix view to arrange
%call graphs in a grid based on similarity using a variety of tree distance
%measures.
%%
%However, not only does the (dis)similarity matrix view suffer from scalability issues, 
%it often requires the user to perform pairwise
%comparisons~\cite{andrews2009visual} by matching individual runs.
%%
%However, traditional tree distance measures can only provide a single value
%to represent how (dis)similar a call graph is from the other, which often can be difficult to interpret the differences.
%%
%\hb{the above is not clear. what do you want to say?}
%In particular, it would benefit in identifying the call site or a particular caller-callee relationship that was responsible for differences between two graphs.
%%
%To this end, we employ one-such technique, called DeltaCon~\cite{koutra2016d}
%which employs spectral graph techniques to quantify the similarities and also
%identify the node or edge of significance.
%
Nevertheless, the more-general task of exploring ensembles of call graphs and
study performance variability is not supported by existing tools.
Instead, experts generally visualize the different call graphs individually to
understand the general calling structure and use simple statistical measures to
understand the distribution of runtimes.

\para{Visual comparison of graphs and trees.}
%To reduce visual complexity and improve readability, graph data is often
%visualized as trees~\cite{ahn2009scalable, isaacs2018preserving}, including the
%case of CCTs and call graphs~\cite{adamoli2010trevis, nguyen2016vipact,
%bergel2017visual, nguyen2019callflow}.
%
%Here, we focus on the works that deal with the comparison of multiple trees for various user tasks; for an exhaustive review of visualization techniques for multiple trees, the reader may refer to 
%the survey by Graham \etal~\cite{graham2010survey}.
%
Comparison is often a key component of data analysis, especially to form
and validate hypotheses~\cite{zhao2012facilitating,
schulz2013design}.
Multiple comparative visualization tools support comparison between two trees
(\eg, the original Unix \texttt{diff} program, or specialized tools such as
TreeJuxtaposer~\cite{munzner2003treejuxtaposer}, Mizbee~\cite{meyer2009mizbee},
Code flows~\cite{telea2008code}).
The challenges in identifying the differences between two trees were summarized
by Graham and Kennedy~\cite{graham2010survey}, and later generalized 
%to the field information visualization 
by Gleicher \etal~\cite{gleicher2011visual},
who provided a taxonomy of visualization designs to support comparison of two
trees.
% In general, two key types of differences exist: a) structural difference
% between two trees, and b) data difference of node attributes.
%
In general, three approaches to comparative designs are \emph{juxtaposition}
(showing different objects separately)~\cite{holten2008visual},
\emph{superposition}~\cite{malik2010comparative} (overlaying multiple objects),
and \emph{explicit encoding}~\cite{amenta2002case, hong2003zoomology} (using an
alternate visual medium to show differences).
%
% Explicit encoding: Visualizing Changes of Hierarchical Data using Treemaps
%

% \hb{the next two paragraphs need work.. not clear why we're talking about
% phylogenetic trees. barcodetree needs more explanation.. what is (T6)?}
% Comparison in multiple trees.
A number of works have focused in comparing multiple trees, especially for
phylogenetic trees~\cite{bremm2011interactive, fu2017ancestral,
liu2019aggregated}.
These works support various analytical tasks using juxtaposed views combined
with user interactions to highlight similarities and differences between two
trees.
However, these techniques cannot scale to a large number of trees and are designed to address the specific needs for phylogenetic trees.
Recently, Barcodetree~\cite{li2019barcodetree} suggest using a bar-code like
visualization for comparing a large number of trees.
However, their approach is primarily suited for stable and shallow trees, and
cannot be used for call graphs, which are significantly deeper.
Focusing on Sankey diagrams in Business Intelligence applications, Vosough
\etal~\cite{Vosough2017} provided a novel approach to visualize uncertainty in
flow diagrams by imposing gradient-based fading on the boundaries nodes of the
Sankey diagram. Although quite powerful, this approach is unable to show the
variability in the distribution of the Sankey flow (\eg inclusive time in call
graph nodes).

%% file: 4_motivation.tex
\section{Domain Problem Characterization}
\label{sec:domain}

Comparing several performance profiles is of significant interest to
application developers aiming to identify performance bugs across versions of a
code or to understand how different application parameters and/or initial
conditions may affect the performance~\cite{patki2019performance,
    chunduri2017run}.
HPC experts are also interested in evaluating ensembles of performance profiles
to assess the role of architecture choices, machine environments, MPI
configurations, \etc, to identify optimal execution and deployment
modes~\cite{wright2009measuring, bhatele2013there}.
Especially with increasing computational capabilities, there is a growing need
for interactively exploring large ensembles of call graphs --- a task not
possible to support using current techniques.

Previously, we developed CallFlow~\cite{nguyen2019callflow} to enable
interactive exploration of call graphs. Although successful for its target
problem, the domain experts frequently reported that CallFlow's inability to
compare multiple call graphs significantly handicapped their analysis.  Indeed,
comparing many graphs was a tedious task due to the need for manually performed
visual comparison via pairwise juxtaposition~\cite[Fig.~8]{nguyen2019callflow},
which not only imposes significant cognitive burden on the user but also poses
obvious scalability challenges.

\para{Requirements.} In this work, we tackle this general problem of
\emph{exploring the performance variability captured by large ensembles of call
    graphs} through a new interactive visual design. Through close collaboration
with domain experts at our institute, we identified four specific requirements
for a new visualization tool.

\para{R1. Compare two call graphs (\ie a \emph{diff} view).} In many cases, the
users are interested in comparing the performance of two executions. Such a
comparison must highlight faster/slower portions of a given execution to enable
easy identification of performance bottlenecks.

\para{R2. Visualize performance variability within several call graphs.}
Working with more than two graphs, users are interested in identifying the
performance variability within a single function or a single module (in
general, a single supernode).  Whereas simpler statistics, such as mean,
median, and variance, are easy to compute already, the users are instead
looking for a visual depiction of the entire distribution to allow them to
understand the overall trend as well as identify outliers.

\para{R3. Visualize additional fine-grained details on demand.} Although a
simplified visual layout is essential for the complete graph, which are
typically very large, additional details, \eg about the module subtree, should
be available to the user on demand.

\para{R4. Interactive exploration and graph operations.} The users desire the
above functionality as well as several other graph filtering and splitting
operations~\cite{nguyen2019callflow} be interactive or near-interactive.

\para{Assumption.} \label{sec:assumption}
We note that different call graphs of the same application may differ in
topology, \eg differences in call paths due to application parameters or MPI
implementations. Here, we assume that such differences are minor (\eg only a
few missing nodes or different edges) and that all ensemble members largely
have the same topology.  Any significant differences (\eg ensembles
representing different applications or completely distinct functionality) are
out of scope for our tool and may require different visual encoding than the
one adopted here.

%% file: 5_visual_interface.tex
\section{Visualization of Ensembles of Call Graphs}

To build an effective tool supporting visual comparisons, we closely follow the
design considerations proposed by Gleicher~\cite{gleicher2017considerations}.
After presenting a detailed description of the call graphs and associated the
data (\autoref{sec:data}), we identify the \emph{targets} for visual
comparative analysis and \emph{actions} that reveal the relationships between
and within these targets (\autoref{sec:targets}) to support the requirements
\textbf{R1--R4}.  Next, we present our \textit{strategies} to scale the
comparative visualization to a larger number of call graphs
(\autoref{sec:strategies}), followed by the description of the visual design
and interactions to support comparison and analysis of ensemble of call graphs
(\autoref{sec:design}).

\input{5_0_data}
\input{5_1_targets}
\input{f2_ensemble_supergraph}
\input{5_3_strategies}

\input{f3_system_overview}
\input{5_4_design}
% \input{5_5_interactions}

%% file: 5_0_data.tex
\subsection{Description of Call Graph Data}
\label{sec:data}

The given sampled profiles are converted into \emph{GraphFrames} using
Hatchet~\cite{bhatele:sc2019}, an open source profile analysis tool.
A GraphFrame~(\gf) is a Hatchet construct that consists of two data structures:
a \textit{directed acyclic graph} (\dg) that represents the CCT or call graph,
and a Pandas~\cite{mckinney2010data} \textit{DataFrame} (\df) that stores the
associated performance metrics.  Hatchet represents collected performance
metrics into individual columns of \df and establishes a consistent indexing
scheme to link the call site in \dg to its performance data in \df, which not
only enables combining data operations with graph operations, but also improves
the scalability of various analytic tasks.

Each call site in \dg is associated with semantic information from the source
code, \eg load module, file name, line number, obtained either automatically
via the profiler or through explicit user annotations.  This semantic data is
extracted from \dg and stored as additional columns in \df for fast access.
Furthermore, the performance metrics recorded on each call site typically
contains metrics from multiple processing units (\eg MPI ranks).
%
% It is important to note that some MPI rank contributes independently to the
% total inclusive runtime of an application. 
%\hb{what does this mean?}
%
Fast access to the data frame is facilitated through hierarchically
multi-indexing \df, where the indices are \emph{module}, \emph{function name},
and \emph{MPI rank}, respectively.  This multi-index representation also
facilitates easy integration of higher-order statistical summaries (\eg
variations across runs, and variations among ranks) over different levels of
details (\eg modules, call sites, or anywhere in between).

In the case of ensembles of profiles, each representing the performance of an
application under different executions, the different \gfi are stored
individually. Generally, meta information about the different executions are
available, \eg machine architecture and/or environment (\ie the number of cores
or processes, maximum available bandwidth, and load capacity of the super
computer), or application parameters.
These \emph{execution parameters} are also stored as additional indexed columns
in the corresponding \dfi.

%% file: 5_1_targets.tex
\subsection{Identification of Targets and Actions}
\label{sec:targets}

\para{Comparison targets} are data entities integral to the comparison task,
\ie the specific data elements being compared.  Here, we identify four
comparison targets.
Whereas the first three targets are \emph{explicit}, \textbf{T4} is an
\emph{implicit} target because it is not known from the data itself, but rather
is based on analysis of the ensemble~\cite{gleicher2017considerations}.  Here,
comparing explicit targets is challenging because the comparison with respect to
call graphs is difficult, and the ensembles pose scalability issues.
On the other hand, the challenge with implicit targets is that they tackle
unknown elements that requires user knowledge to interpret the differences.

\para{T1. Calling contexts} are an overarching target for our comparative
visualization.  Any optimization efforts usually focus on identifying parts of
the applications that must be improved, for which, understanding the entire calling
context of the application is essential to visualize.  Furthermore, different
execution modes (\eg MPI libraries and application parameters) may lead to
slight differences in the calling contexts.  Understanding such differences is
also important to reason about the variability in the performance within
ensembles.

\para{T2. Performance variability across runs.} Production codes with high
run-to-run variability may cause unpredictable slowdowns and diminish the
reproducibility  of successful experiments.  Highlighting such variability, at
both module (summarized) and call site (detailed) levels, is one of the most
critical comparison targets in this work.

\para{T3. Performance variability across MPI ranks.} For MPI-enabled
applications, the overall performance degrades when the runtime is less uniform
across resources~\cite{pearce2012quantifying}, \eg due to load imbalance.
Identifying heavily over- or under-utilized resources from the runtime
distribution is important to design codes that can achieve peak performance.

\para{T4. Execution parameters.} In many cases, an application's performance
may be correlated with certain execution parameters~\cite{wright2009measuring}.
Therefore, identifying the parameters with more significance can help
understand explored multiple optimization strategies simultaneously.

% -------------------------------------------------------------

\para{Comparison actions} are visualization tasks needed to understand the
relationships between and within comparison targets.
Formulation of actions influences the visual encoding and the user interactions
that link individual visual components. Visualization of ensembles of call
graphs broadly require six actions.

\para{A1. Compare calling contexts across runs.} Due to potential differences
in calling contexts of the different executions (\textbf{T1}), it is essential
to highlight any differences, \ie missing nodes/edges in the graph.

\para{A2. Compare performance variability across runs.} Displaying the
aggregated performance from individual executions is not a scalable approach
considering the effort from the user, and the corresponding visualization
complexity.  Instead, the recorded performance (\textbf{T1}) and calling
contexts (\textbf{T2}) must be summarized not only within a single execution,
but also across multiple call graphs.

\para{A3. Compare performance variability across MPI ranks.} Summary statistics
(\eg mean runtime) across MPI ranks (\textbf{T3}) is not sufficient to analyze
runtime distribution across resources because the distribution is not expected
to be normal.  To aid the exploration, complete distributions are essential to
visualize.

\para{A4. Compare call graphs across levels of detail.} Although comparisons
across super graphs (semantically aggregated call graphs) is useful, when
experts typically identify problem areas, they are interested in looking at
selected regions in more detail. Therefore, it is important to manage
fine-level details and visualize them upon request.

\para{A5. Compare two call graphs.} Differences between two call graphs can
exist structurally (\textbf{T1}), as runtime variations (\textbf{T2, T3}),
and/or in execution parameters (\textbf{T4}).  Simple, visual representations
of such differences is of value to the user.
%but can be visually overwhelming and computationally expensive for comparison.

\para{A6. Compare a selected run with an ensemble.} Comparison tasks often
require a {baseline} to compare against.  For all of our targets
(\textbf{T1--T4}), it is important to compare a selected run with the ensemble
behavior.  An important constraint to consider is that the constructed baseline
must match the expert's understanding of the overall calling context, despite
minor differences in the individual ensemble members.

%% file: f2_ensemble_supergraph.tex
\begin{figure*}[!t]
\vspace{-0.5em}
    \centering
    \includegraphics[width=0.82\textwidth]{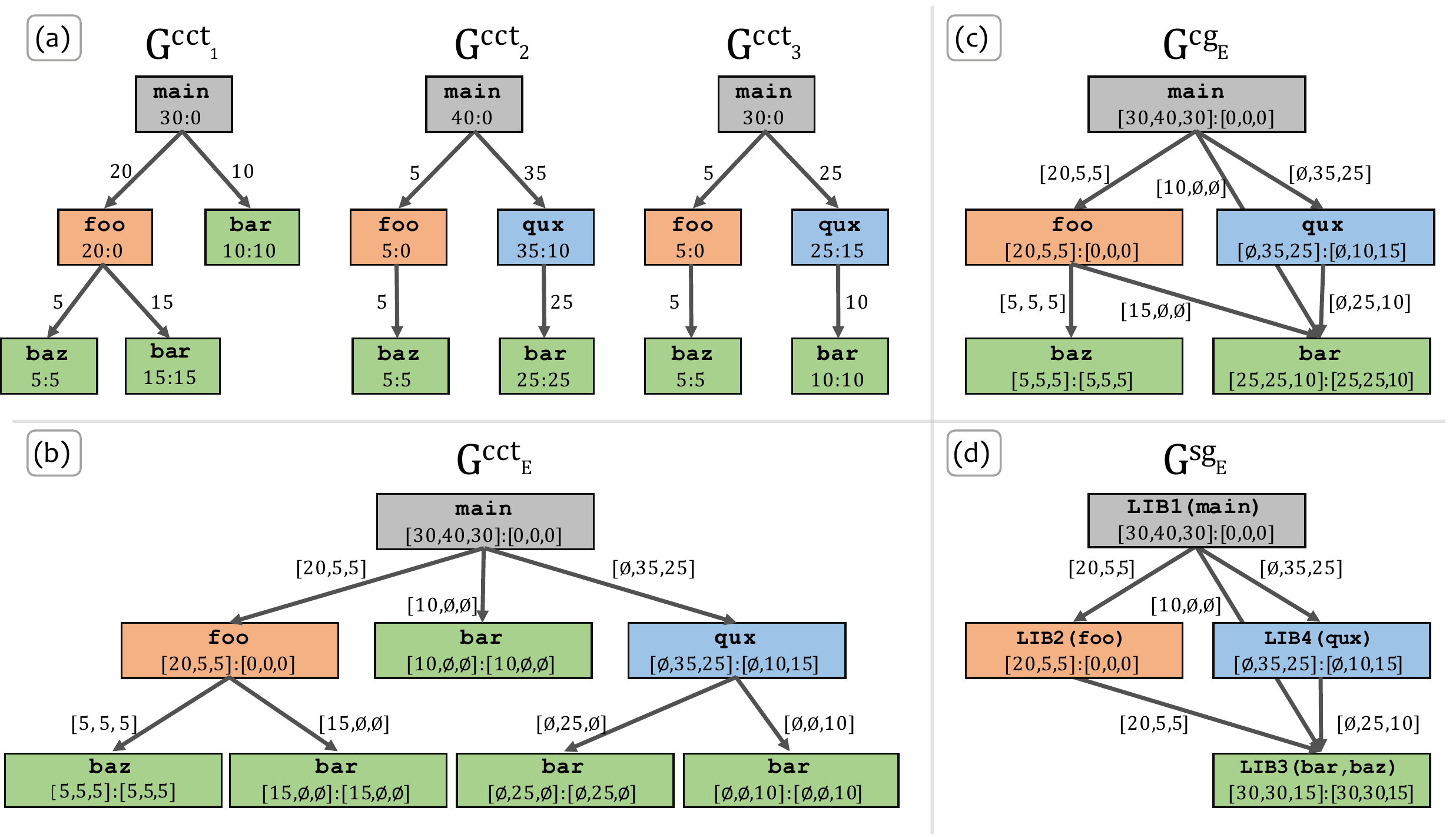}
    \caption{%
        Construction of \textit{ensemble super graph} from 3 given graph
        frames shown in (a).
        (b) First, the ensemble CCT, \dgce, is constructed to include all unique
        calling contexts across \dgci. The performance metrics for corresponding call
        sites (nodes in the graph) are concatenated into associated vectors, where
        missing nodes in any given CCT are denoted with $\varnothing$ in the vector.
        (c) Next, \dgce is converted into an ensemble call graph, \dgge, by grouping
        call sites representing the same function into a single node. The associated
        metric vectors are element-wise added.
        (d) Finally, semantic information, \eg library names, are used to group call
        sites (nodes in \dgce) to create the ensemble super graph.\vspace{-1.8em}}
    \label{fig:ensemble_supergraph}
    \vspace{-1.5em}
\end{figure*}

%% file: 5_3_strategies.tex
\vspace{-0.5em}
\subsection{Development of Strategies}
\label{sec:strategies}

% \hb{1. It seems that ‘aggregation of call graphs into ensemble graph’ would be
%     one strategy (your 2, 3) 2. Then 4, and 5. And then 1 3. You also want to
%     write about detailed information. That your strategy is to show the user
%     module/call site level information using alternate ways because you do not
%     want to change the overview (so you you use complementary view)}

The scale of sampled profiles collected by HPC experts can vary from tens to
hundreds depending on the experiment.
Hence, it becomes vital for CallFlow as a visual analytic tool to handle
scalability arising for three reasons: 1) number of sampled profiles, 2) number
of call sites in each profile, and 3) number of ranks associated with each call
site in each run.
In this section, we detail the strategies we adopt to support scalable analysis
of a reasonably-sized collection of profiles.
%
% Our efforts can be split in two directions: a) support scalable data processing
% for the targets, and b) design appropriate visual encodings that aid in
% performing the listed actions(\textbf{A1-A6}).

% summarizing and comparison task.
%
% For more discussion on performance, refer to the discussion (\autoref{sec:performance}).

\vspace{-0.2em}
\subsubsection{Construction of Ensemble GraphFrames}
GraphFrames (\gfi) are the data source for CallFlow, by feeding appropriate
information about the \textit{comparison targets} to the visual interface upon request.
%
% and responds to different requests from the visual interface through user
% interactions.
Furthermore, different \textit{comparison actions} require different data
operations (\eg statistical summaries of data), and/or the graph operations (\eg
grouping, filtering, matching) to be performed before visualizing the results.
These operations demand a consistent and scalable approach to store, access, and
modify the \gfi.
%
% To facilitate comparison of the calling contexts (\textbf{A1}) and associated
% performance variability(\textbf{A2}) across multiple runs, both the data
% structures, dataframe(\dfi) and graph(\dgi) must be stored and accessed in a
% concise manner.
%
Here, we describe how to concisely represent the entire ensemble as a single
\emph{ensemble GraphFrame} (see \autoref{fig:ensemble_supergraph}(d))
that supports all required queries.

% the individual
% GraphFrames (\gfi) must be accessible.

\para{Step 1. Unify all DataFrames.} First, we perform the \textit{unify}
operation to concatenate \dfi's into an \textit{ensemble dataframe} (\dfe).
This operation is similar to Hatchet's unify implementation, except for the
reindexing step, where we add column \textit{``runName''} to tag the call sites
belonging to individual runs.
Consequently, the \textit{``runName''} column becomes the primary index of the
multi-indexed hierarchy of \dfe.
However, for multiple large call graphs (say 100's of call sites in each
profile), \dfe can quickly becomes data-intensive and eventually costing time
and memory for various data operations.

To improve the processing time for multiple data operations operation, we
abstract the information stored in \dfe at two levels of data
coarseness (\textbf{A4}), namely \textit{module-level} and
\textit{name-level}.
The module-level granularity groups the data points based on the library (\ie
module column), while the node granularity groups by the call site's name (\ie
name column).
Both granularities of \dfe are stored as separate files using the
HDF5~\cite{hdf} data model, which is community standard
format used in many scientific applications for
storing hierarchical-structured datasets.
Additionally, HDF5 allows us to attach semantic information (\ie library name
and call site name) as \textit{``attributes''} that group data points together
for faster data lookup.

% One key feature of HDF5 is that attributes can be attached to group data points
% together and this grouping facilitates faster data lookup.
%
% To exploit this feature, we use the semantic information (\ie library name, and call site name) to group the
%
% Since most operations occur at the module-level information or

% The ensemble dataframe is grouped into chunks by \emph{module},
% \emph{callsite}, \emph{rank}, depending on the level-of-detail and the
% aggregation required by operations and are stored using the HDF5 file format.
%

% \subsubsection{Construction of the Ensemble Super Graph}

\para{Step 2. Unify all calling contexts.}
Next, we perform the \textit{union} operation on the \dgi's to construct an
\textit{ensemble CCT} (\dgce).
The union operation aggregates the performance metrics for each callsite and
merges the calling contexts that share the same caller-callee relationship
across \dgi's.
A call site from \dgi is considered equivalent to a call site from \dgj if their
complete calling context is same across the two runs.
The aggregated performance metrics corresponding to each callsite is stored as
vectors.
In \autoref{fig:ensemble_supergraph}(a), we demonstrate the union operation
using three sample \dgi's comprising of 5 call sites that belong to 4 libraries.
Although there are minor differences in the graph (\eg missing nodes),
the union operation accounts for such inconsistencies
by assigning a null value ($\varnothing$) to the missing nodes in any \dgi.
For example, whereas the calling contexts of \pcode{baz} remains consistent
across the three CCTs, the
contexts of \pcode{bar} and \pcode{qux} differ across the runs.
The resulting \dgce (see \autoref{fig:ensemble_supergraph}(b)) represents a
super set of the given \dgi's, preserves all
calling contexts, and merges the nodes with identical contexts.

Next, \dgce is converted into a
\textit{ensemble call graph} (\dgge), according to the usual definition,
\ie call sites with same function name are merged
(see \autoref{fig:ensemble_supergraph}(c)).
The associated vectors are element-wise added to summarize the performance
runtime on each callsite.
Finally, the call sites are filtered
(especially for large call graphs) by inclusive/exclusive runtime
using user-defined thresholds, and later grouped with the available semantic information, to
construct CallFlow's module-level super graph,
which we call \textit{ensemble super graph} (\dgse).

The final construct, the \textit{ensemble GraphFrame} (\gfse) is then
created by combining \dfe with \dgse.
The ensemble GraphFrame not only captures all equivalence relationships
among call graphs
using nodes and the edges store the subtle differences in the performance and
calling structure.
The aggregated \dgse can now be used as the \textit{baseline} super graph to
compare against the ensemble (\textbf{A3}).

Both steps, the unification of DataFrames and the unification of calling contexts, are
done as a preprocessing step, to provide the user with options to filter large call graphs
and modify the semantic information, if necessary (as done in the original CallFlow~
\cite{nguyen2019callflow}).
To aid with filtering and grouping operations, the users still
have an option to load a single profile using CallFlow and verify the resulting
module-level super graph before studying the ensemble.

\subsubsection{Enriching Targets into Ensemble GraphFrame}

Although, \gfse succinctly represents an ensemble of sampled
profiles, it does not capture the potentially critical patterns hidden among the
comparison targets.
%
% that relate each target to the
%
Therefore, the next challenge is to develop strategies to help visualize summaries and enable comparison among targets.
%
%
% One fundamental difference between summarization and clustering is that the
% former finds coherent sets of nodes with similar connectivity patterns to the
% rest of the graph, while clustering results in coherent, densely connected
% groups of nodes.

% \para{Mean runtime}
\para{Summarize performance metrics across runs.}
Runtime metrics (\ie inclusive/exclusive runtime) are valuable indicators of
performance slowdown among call sites.
For each node in \dgse, aggregated runtime metrics are stored as a
$n$-vector, where $n$ is the number of runs in the ensemble.
In the more-common scenario of parallel codes, the metrics are
first aggregated across threads/processes/MPI ranks. Such aggregation
is important to be able to compare profiles with different number of ranks,
\eg 8 vs.\ 512, consistently. In particular, to summarize
the runtime metrics across multiple
executions (\textbf{T2}), we take \textit{mean} runtime across all ranks in
an execution to be the representative runtime for the corresponding run.
%
%2-dimensional \textit{vector} with dimensions,
%$n \times m$, where $n$ represents the number of runs in the ensemble, and $m$
%is the number of MPI ranks in a particular execution.
%%
%To summarize the runtime metrics across multiple executions (\textbf{T2}), we take the mean runtime across the ranks.
%%
%Mean runtime as a metric summarizes the distribution of process runtime even for runs with varying MPI ranks.
%
Given a single value of the metric per node per execution in \dgse, we next capture the
distribution of the metric across executions for each node to create the
\textit{ensemble distribution}.
Here, we compute histograms for each node, the range of which spans minimum to
maximum of the runtime for any execution.

%However, comparing mean runtimes across call sites does not provide information
%regarding the underlying trend among the executions.
%%
%To capture the trend, we bin the mean runtimes across processes to create 
%the \textit{ensemble distribution}.
%%
%Our binning strategy sets the range of bins between minimum and maximum of the
%runtime of the callsite across runs.
%%
%Finally, we also calculate the bin of the mean runtime of all the  runs in the
%ensemble distribution, allowing us to compare the target run's distribution against the ensemble.

\para{Summarize performance metric across MPI ranks}
Although, the ensemble distribution conveys the trend of the mean runtime per node, it is equally important to study the runtime distribution across MPI ranks (\textbf{T3}).
To this end, we preserve the per-rank data on the server, and supply to
the visualization upon demand. Such data can be visualized in two ways.
We show a box-plot visualization for the run time across MPI ranks
for each call site. Box plots are a powerful way to capture the limits, means/medians,
and standard deviations/quartiles, as well as outliers in the data.
We also show a histogram of the run times per rank for the entire ensemble
as well as for a selected target run.

%%
%The minimum and maximum runtime reveals the lower and upper bound of the distribution.
%%
%However for the comparing runtime, mean runtime and standard deviation is a more
%useful statistic since mean provides a bigger picture of how the runtimes are
%distributed, and standard deviation.
%%
%However, both mean and standard deviation as a statistical measures are not
%sensitive to outliers, which in our case causes load imbalances.

%Therefore to study the process level distribution, we calculate quartiles to
%observe and explore the variation in the observed runtime distribution for each
%callsite.
%%
%Quartiles separate the MPI-rank runtime distribution into four equal-sized
%groups.
%\sk{Need to explain the quartiles a bit more. along with the outliers.}

\para{Reveal the subtree within a module.}
The ensemble super graph \gfse  is designed to provide a module-level
overview (by default). Upon request, the subtree inside a selected module
can be extracted revealing further details about the ensemble. For such cases,
%%
%However, in most scenarios, HPC experts are intersted in drilling down the call
%sites inside to identifying particular call sites that cause performance
%behavior.
%
Previous version of CallFlow~\cite{nguyen2019callflow} used graph splitting
operations to reveal call sites of importance, but this demands the user to sort by the
exclusive runtimes and then reveal a selected call sites.
However,
for an ensemble of runs, this strategy fails because it requires the user
to compare exclusive runtime from $n$ runs, making it challenging to
analyze the trend in the distribution, which is critical for the identification
task.
Instead, it is important  to reveal all call sites belonging to a supernode on user's demand.
To this end, we perform a breadth-first traversal on the ensemble CCT
(\gfce) from the entry function of all the supernodes and construct a hierarchy tree, called \emph{supernode hierarchy}.
% ($H_m$).
%
Using \gfce ensures that \emph{all} call sites present in \emph{any} call graph
are accounted for.
Here, we use the $\varnothing$ from the aggregated runtime metric vector to
determine if a particular call site exists in a selected run.
To ensure the calling context inside the module to be a tree, we break all the
cycles inside the supernode (as already done in CallFlow~\cite{nguyen2019callflow}).
%
%Finally, since the calling contexts can slightly vary between runs, it is important to 
%know if all call sites are actually present in the $H_m$.
%%
%However, comparing calling contexts across super graphs is a computational heavy
%because sub-graph matching between graphs in an ensemble is a NP-hard
%optimization problem.
%%
%Instead, 

%
% in learnt from experts that they are interest to represent the mean of the inclusive runtimes across processes.
% %
% % why mean?
% This is primarily because it represents the
% Additionally, This is primarily because the runtimes of processes can be very skewed where most values are 0 and this could drastically affect the perception of the user.
% %
% Although, multiple processes with zero runtime signifies a skewed load balancing which could cause bottlenecks, the experts consider it to be a detail-on-demand information.
% %

%
% The kernel density estimation approach overcomes the discreteness of the
% histogram approaches by centering a smooth kernel function at each data point
% then summing to get a density estimate.
% %
% However, KDE shows the multi-modality of this bivariate distribution, which
% contains multiple bumps that cannot be captured easily by any parametric
% distribution.
% %

%% file: f3_system_overview.tex
\begin{figure*}[!t]
\centering
\vspace{-0.5em}
\fbox{\includegraphics[width=0.9\textwidth]{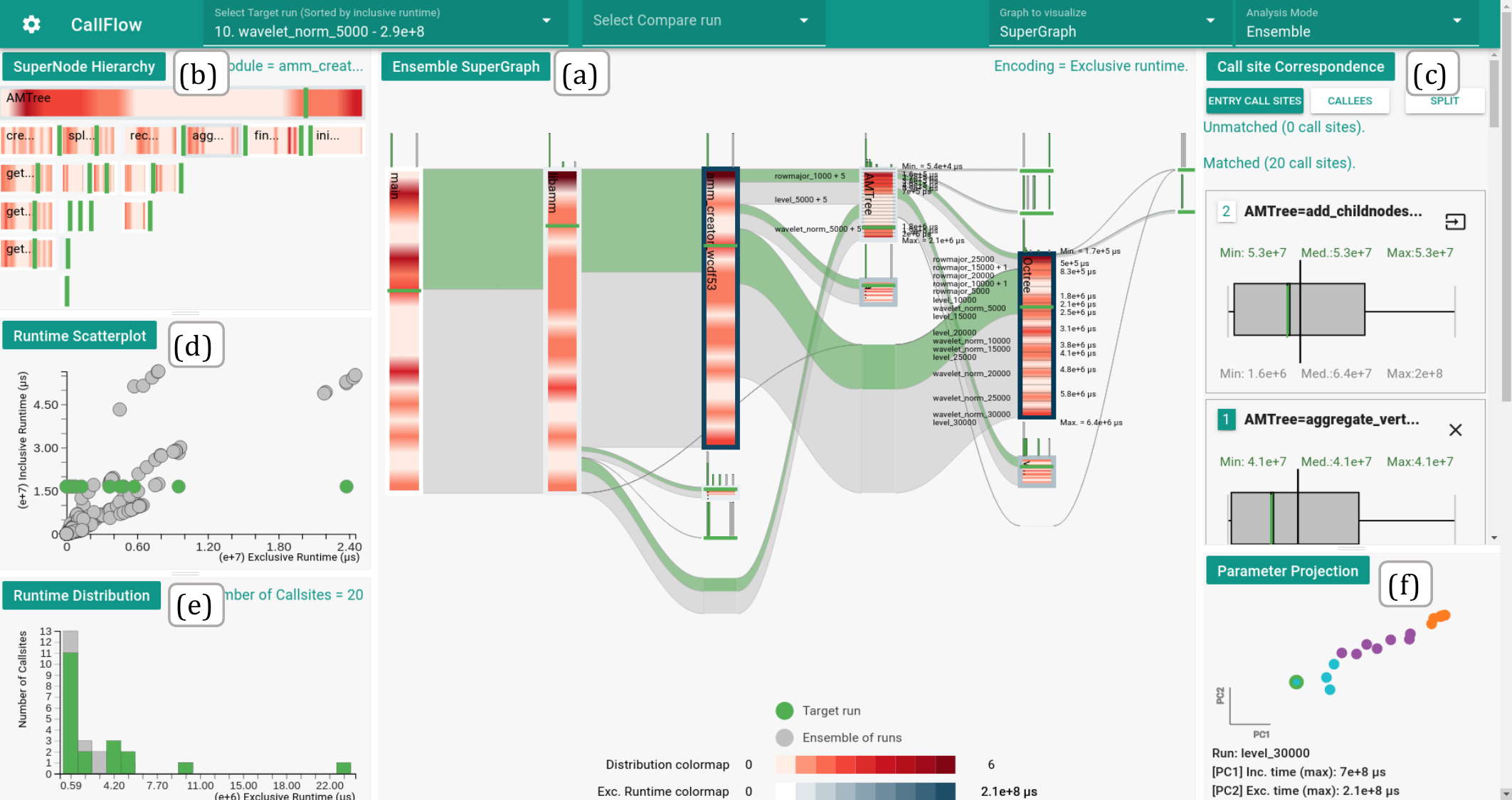}}
\caption{CallFlow enables scalable visual analytics of ensembles of call graphs 
using a novel visual design, \emph{ensemble-Sankey} (a), and several linked 
views (b--f), which provide additional, fine-detail information (\eg hierarchy within 
a node of the Sankey), as well as statistical descriptions of the data (\eg run time distributions). 
Through interactions with visual elements as well as UI-based options, the tool allows 
thorough exploration of the ensemble data, serving a wide variety of application-specific 
requirements. 
\label{fig:system-overview} \label{fig:amm}\vspace{-1.6em}}
\vspace{-0.5em}
\end{figure*}

%% file: 5_4_design.tex
\subsection{Design of Visual Analytics System}
\label{sec:design}

As shown in~\autoref{fig:system-overview}, the visual interface of
(the new version of) CallFlow comprises of five
inter-linked views and three analytic modes that form a comprehensive visual interface
to support the comparison actions. The source code is available in~\cite{sourcecode}.

%
% The three analytic modes are, a) ensemble mode to summarize an ensemble of runs,
% b) target-ensemble mode to compare a selected run against the ensemble, and c)
% target-target mode to detect runtime differences.
%
% visual encodings that help summarize in the ensemble mode, compare in the
% ensemble-target mode and highlight differences in the target-target mode.
% 
%Below we describe the view and its functionalities.

\subsubsection{Ensemble-Sankey: The Ensemble SuperGraph View}
Previously, CallFlow~\cite{nguyen2019callflow} utilized the {Sankey}
layout to visualize a (single) super graph. In this work, we expand this visual
design to support visualization of ensembles of super graphs, and create a
novel design -- the \textit{ensemble-Sankey}.
%
% Ensemble Super Graph view, (\autoref{fig:system-overview}(a)), visualizes the
% constructed \gfse using a Sankey Diagram similar to the \textit{single-Sankey},
% which visualizes the resource flow in a single super graph.
%
Our decision to preserve and enhance the Sankey layout emanates
from the effectiveness of the existing decision.
In particular, the resource flow conveyed by Sankey layout provides an excellent way to
represent calling contexts and associated performance metrics, as it matches
experts' intuition about the internal representation of the application flow.
%
%a different 
%design all together could affect the perceptual ability of the user~\cite{szafir2016four}, and b)
%the resource flow conveyed by Sankey diagram matches the internal representation
%of the application flow for the HPC experts.
% The decision to stick with the Sankey diagram representation is built on the fact that it matches internal understanding for the HPC experts.
%
Furthermore, we assert that perceptual simplicity is vital for the comparison task,
as the user has to build the proficiency to ``spot the differences'' in an
ensemble of runs.
However, using the existing encoding of nodes and edges are not directly
usable for the ensemble, since only a single graph may be visualized.
Instead, we adapt node and edge encodings to concisely represent an
ensemble in the proposed \emph{ensemble-Sankey} design.

% --------
\para{Ensemble Nodes.} In the (standard) Sankey,
each supernode is visualized as a rectangular bar, whose height corresponds to
the ``resource of interest''. For the application at hand, the resource is the total
time spent within this module and all its callees (\ie the sum of the inclusive runtime
of all its entry functions)~\cite{nguyen2019callflow}.
In the case of ensembles, however, there are many graphs with different
values of runtime. Nevertheless, continuing the same interpretation of
``resource flow'', it makes sense to aggregate the runtimes across
all executions. We therefore scale the height of a given supernode in
the ensemble-Sankey proportional to the \emph{maximum} inclusive metric across all executions
because the goal is to show the entire distribution (across executions) onto a
single visual element of consistent shape and size. Under such encoding,
the ``standard supernode'' of any given execution simply becomes a subset
of the ``ensemble supernode''.
The width of the rectangle is computed as before~\cite{nguyen2019callflow}.

With the inclusive metric mapped to the height of the rectangle, we are also
interested in highlighting the exclusive metric. This is achieved by drawing
colormapped borders of the node. As above, we use the \emph{maximum}
exclusive metric across all executions, as it can help identify exceptionally
slow runs.
Although the default border coloring is the (maximum) exclusive metric, the user
can interactively avail other options, such as color by (maximum) inclusive
metric.

%Instead of encoding the entire width of the rectangular bar with a single
%\textit{runtime metric}, we use the borders for the same.
%%
%The runtime borders signify the maximum contribution to the runtime among all runs in the ensemble. 
%%
%By default, we choose the runtime metric as exclusive runtime since it is 
%useful for identifying call sites that cause performance slowdowns.
%%
%%
%The user can change the interested runtime metric and color map for the runtime
%distribution from the dropdown settings icon displayed by clicking the icon in
%the top left\lcircle{1}.
%%
%The corresponding colormaps are shown below the ensemble-Sankey\lcircle{2}.

We add borders to supernodes to make room for additional
information inside the rectangle. Previously, CallFlow simply colored the entire node,
which we will now use to show the distribution of a chosen metric.
Although the distribution of inclusive metrics is the most meaningful choice
(since the height maps to maximum inclusive runtime), the user may also use
an exclusive metric (as shown in~\autoref{fig:system-overview}).
In particular, we compute a histogram of the chosen metric and map it
vertically to the height of the supernode.
Since the true histogram may contain sharp spikes (we do not necessarily expect to
find a smooth distribution), we use smooth gradient fading to ensure all peaks are
highlighted.
Although we could directly compute a density estimate (\eg using KDE)
instead of a histogram to generate a smoother distribution, such techniques
introduce additional parameters (\ie kernel widths), which are typically harder to interpret
and may miss features regardless due to an unsuitable choice.
Instead, we use histograms (with a customizable bin count) and use linear gradients
to smooth the distribution.
The resulting \textit{ensemble gradient} therefore shows the full distribution
of the selected metric across runs.
The default choice of color for the distribution is a single-hue white--red colormap.
The colors are mapped consistently across all nodes to allow evaluation of distributions not just
within a single supernode, but also across supernodes
(\textbf{A2}).
%

%Next, the ensemble distribution is overlaid along the
%height of the node using a linear \textit{ensemble gradient}, where the top and bottom of the node corresponds to the minimum
%and maximum of a selected runtime metric across the runs, respectively.
%%
%Gradients employ a focus-based technique to reveal the number of executions
%belonging to a particular bin by applying blurriness around the bin.
%%
%By default, we use a linear white-red color map to color the gradients and set the number of bins for ensemble distribution to be 20.
%%
%Encoding the distribution on directly on the node directs the user to observe
%the selected runtime metric's trends across the runs.
%
%Using a colored gradient can help the viewer an insight into the runtime variations (\textbf{A2}) by noticing how ensemble gradients either spread (high variation) or shrink (low variation).
%

Additional runtime information can be revealed by toggling the
\textit{text-guides}, which not only display the minimum, and maximum runtimes of
the executions among the ensemble, but also mark the bins of the histogram.
% corresponding bins using gray
%lines that split the ensemble distribution into equal parts execution's.
%
By default, the text-guides reveal only the number of executions in the corresponding bin,
but additional information such as ``runName'' can be examined by click on the text-guides.
For example, in \autoref{fig:system-overview}, the gradients of the \pcode{AMTree} supernode
show the mean exclusive runtime skewed towards the two extremes.

\para{Ensemble Edges.}
As in the (standard) Sankey, superedges (which connect two supernodes)
encode the flow of the inclusive metrics in \dgse. However, in the (standard)
Sankey, the thickness
(in vertical direction) is proportional to the inclusive metric consumed by the target node
(\ie the resource being transferred to a target node).
%
% Since the calling contexts can be different among runs, we encode the number of runs using the opacity of the edge.
%
However, this property cannot be enforced for the superedges of an ensemble-Sankey
because of the aggregation (max) performed for each supernode. In particular, since the
maximum values of child nodes may be from different runs, preventing the sum of
all outgoing resources to be equal to the height of the node (minus the exclusive time).
As such, there is no single
unique mapping that can be defined for the height of the superedge.
For example, consider \autoref{fig:ensemble_supergraph}(d), where the
aggregated metrics for each supernode is the maximum of the
corresponding vector, implying the values of the nodes corresponding
to \pcode{LIB1}, \pcode {LIB2}, \pcode {LIB3}, and \pcode {LIB4} to be
40, 20, 30, and 35, respectively. However, the aggregated (max) metrics
for the outgoing edges from \pcode{LIB1} are 20, 10, and 35, respectively,
which sum to 65.
Such scenarios are common in profiles with high performance variability from
different libraries.
To present a consistent encoding and interpretation of superedges, we instead
scale the two ends of a superedge proportional to the respective nodes, thereby,
smoothly varying the height of the superedge (from left to right).
In addition to a neat visualization, this encoding facilitates visualizing a superedge
with respect to both its source and destination.

%
%if we add up the runtime contributions from 
%all the entry functions to a supernode does not necessarily add up to its net inclusive 
%runtime.
%%
%As an example, in \autoref{fig:ensemble_supergraph}, if we add the edge contributions flowing out of \pcode{main} function, the total is 65.
%However, the height of the node is 40. 
%%
%This phenomenon is common when we are studying profiles with high performance variability from different libraries causing the thickness for supered to over flow over the node. 
%%
%This may add additional difficulties to the user perception hindering the user from understanding the application flow. 
%To prevent the above behavior, we rescale the edge based on their out-flow in our Sankey layout optimization algorithm to match the net-flow of a supernode to its height.
%%
%% e some of the executions might may contain multiple supernode that have the maximum inclusive metric across different executions (\ie due to performance variability).
%%
%%
%%
%%
%% Since we consider the flow as the maximum inclusive metric, the 
%%
%% This primarily affects the assignment of thickness for the superedge, as it may overflow at the source node, which could confuse the user's perception.
%%

\vspace{-0.4em}
\subsubsection{Supernode Hierarchy View}
\label{sec:modulehierarchy}

While scanning the ensemble gradients, the user is often limited to only exploring the ensemble distribution for the revealed supernodes.
Using graph operations, such as splitting, can reveal additional interesting nodes but
the graph operation changes the overall context.
For example, splitting a supernode by its entry functions would reveal all the entry functions
belonging to a supernode, but if there are multiple entry functions, the Sankey layout
may change significantly.
%
%This could also jeopardize the user from understanding the resource flow, as  revealed nodes could occupy an entire level in the Sankey layout. 
%
Through discussions with potential users, we instead choose to visualize the additional
details separately, in particular, as a \textit{supernode hierarchy} using an icicle plot, similar to
the flamegraph visualizations~\cite{gregg2016flame}.

Icicle plot places the call sites of the selected library based on their depth inside the supernode
hierarchy from top to bottom.
Each call site in the supernode hierarchy is visualized as horizontal rectangular bars.
As with the ensemble-Sankey, ensemble gradients and runtime borders are used to
encode the ensemble distribution and the runtime distribution for the call site, respectively.
Additionally, when a target run is selected, the supernode hierarchy view also enables the comparison of multiple calling
contexts (\textbf{A1}), as the nodes with $\varnothing$ have no
ensemble gradients that fill the rectangular bar, allowing the user to identify the missing
call sites easily.
Furthermore, call sites of interest from the supernode hierarchy can be selected by clicking
to reveal all call sites in the module's context.
Revealing a call site's calling context in the ensemble-Sankey helps compare the runtimes to
identify performance slowdowns among call sites (see~\autoref{sec:lulesh-study}).
% \hb{last half para is quite unclear}

\vspace{-0.4em}
\subsubsection{Call Site Correspondence View}
\vspace{-0.1em}
Call site correspondence view (see \autoref{fig:system-overview}(c)) lets us focus on
process-level distribution, and also scan call site information based on data and graph properties.
Previously, (single) Sankey's histogram view was used to explore the connection between
slowdowns in MPI ranks and the physical domains.
However, such cases are very specific to the application in study, and do not add any significant
value to the comparison task, which has a broader scope of exploring variations across
multiple objectives.
Additionally, such histograms would only work where all ensemble members
have the same number of MPI ranks.

Instead, we resort to more a traditional visualization --- a boxplot to explore the variation
in the observed runtime distribution for each call site (\textbf{A3}).
\emph{Boxplots} use quartiles of the distribution to indicate the spread of data, and
can also reveal outliers.
%
%Quartiles separate the process runtime distribution into four equal-sized groups
%
In this view, we enumerate the call sites that are not present in the ensemble view, and
highlight the median, the interquartile range (IQR = Q3$-$Q1), as well as outliers
(above and below 1.5$\times$ the IQR.
Additional information is also provided as text labels (\eg minimum, median, and maximum
runtime).
\vspace{-0.4em}
\subsubsection{Complementary Views}
\vspace{-0.1em}
\para{Metric correlation view.}
In order to compare inclusive and exclusive metrics for a given call site, we produce a
scatterplot that captures the correlation between the two (\autoref{fig:system-overview}(c)).
Previously, a similar scatter plot was used in single profile analysis to identify
expensive call sites within a supernode.
For ensembles, each ``dot'' in the scatter plot represents a callsite for a single run, making it
possible to study such correlations across ensemble members, \eg by comparing a
target run to the ensemble, as shown in the figure.
The scatter plot itself is also interactive, as hovering over a dot highlights call site by their names
for the user to compare the runtime metrics(\textbf{A1}).

\para{Runtime distribution view.}
To assess the distribution of runtime, we show histograms for the chosen metric
(inclusive or exclusive) \emph{for a  selected supernode} in three modes:
(1) \emph{call site mode}, which shows the distribution of runtime metric with
respect to individual call sites (\ie the vertical axis is the number of call sites with a
given range (bin) of runtime, as shown in \autoref{fig:system-overview}(e));
(2) \emph{call graph mode}, which shows the distribution with respect to
the call graphs (\ie the vertical axis is the number of call graphs with
a given range (bin) of runtime); and
(3) \emph{MPI rank mode}, which shows the distribution with respect to
the MPI ranks (\ie the vertical axis is the number of MPI ranks with
a given range (bin) of runtime).
These distributions are important to explore for a complete understanding
of the chosen supernode in an ensemble, with or without a target run.

\para{Parameter projection view.}
We also include a parameter projection view (see \autoref{fig:system-overview}(f)),
which is an experimental feature to explore the space of parameters that generate the ensemble.
Here, we normalize all known execution parameters using min-max scaling (\ie normalized to
range [0,1]), all non-numerical parameters are first mapped to integers.
The normalized parameters, along with the exclusive and inclusive runtimes,
are then projected to a 2-D space using Multidimensional Scaling(MDS)~\cite{kruskal1964multidimensional}and clustered using k-means.
The intent is to allow the user to explore potential similarities in the execution
parameters as well as provide a tool to subselect the ensemble --- a task
enabled using a lasso in the projection plot.
However, an important limitation is the interpretability of such projection, especially
due to the string to numerical mapping in certain parameters.
Nevertheless, with appropriate caution in mind, the user can use this tool to further
explore the data.

\subsubsection{Visual Analytic Modes}

\para{Ensemble summary mode}
is the default analytic mode for CallFlow when studying an ensemble of runs
(see \autoref{fig:lulesh-summary}).

\para{Target-ensemble comparison mode}
is triggered when the user selects a particular execution
from the ensemble to study in detail from ``Select Target run'', which lists all ensemble
members.
This operation allows the user to compare a selected run's performance with the
ensemble(\textbf{A6}) across all 6 views.
Then, the visual elements belonging to the selected target run are revealed
    {in green} through out the system (in \autoref{fig:system-overview}) --
for consistent context, the elements corresponding to the ensemble are
shown {in gray}.
In particular, the \textit{target superedges} are overlaid on top of the ensemble edges to compare the proportion of inclusive runtime that calling contexts (\textbf{T1}).
The \textit{target-lines} (\ie green colored lines) place the selected run in the
distribution for each supernodes. (\textbf{T2})
Boxplots are revealed on top of the ensemble boxplot to compare the process runtimes
of the target run vs. ensemble.

\para{Target-target difference mode.}
During comparative analysis, the user might find two executions that exhibit a variation in the
ensemble distribution, or appear an outlier from the projection view.
To allow the user study the differences between two supergraphs, we compute a diff call graph that
subtracts the mean runtime across supernodes.
The resulting \textit{diff view} is visualized as a Sankey diagram and is colored with a green-red colormap, where hues of red color highlight the performance slowdown and hues of green highlight performance speedup.
% and might want to compare theruns with respect to that run.

%% file: f4_lulesh-compare.tex
\begin{figure*}[!t]
    %\captionsetup{farskip=0pt}% <--- no gap at the top
    \centering
    \subfloat[Target-Ensemble view]{
        \includegraphics[height=5.8cm, keepaspectratio]{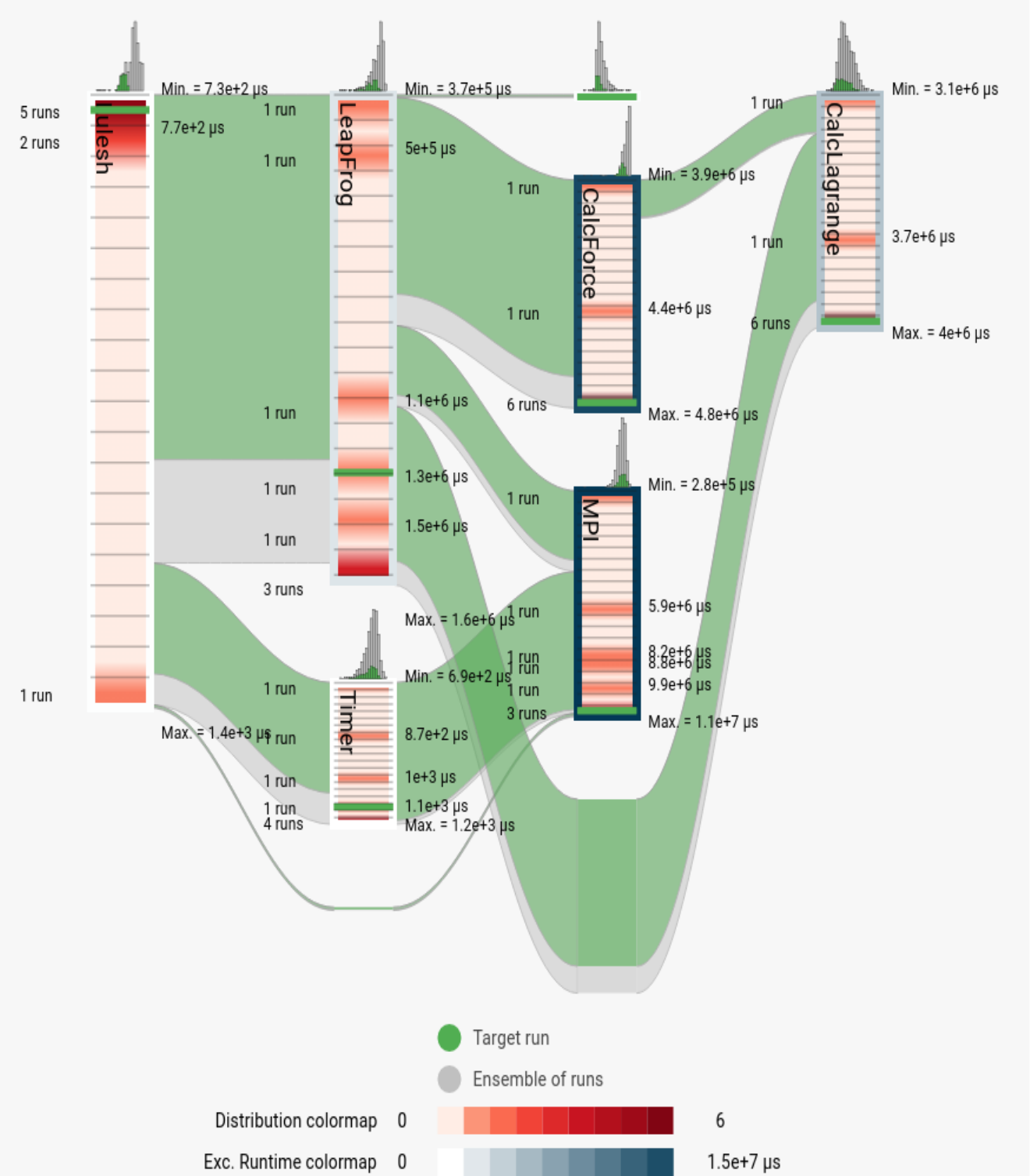}
        \label{fig:lulesh-ensemble}
    }
    \hfill
    \subfloat[Diff view (64-cores $-$ 27-cores)]{
        \includegraphics[height=5.8cm, keepaspectratio]{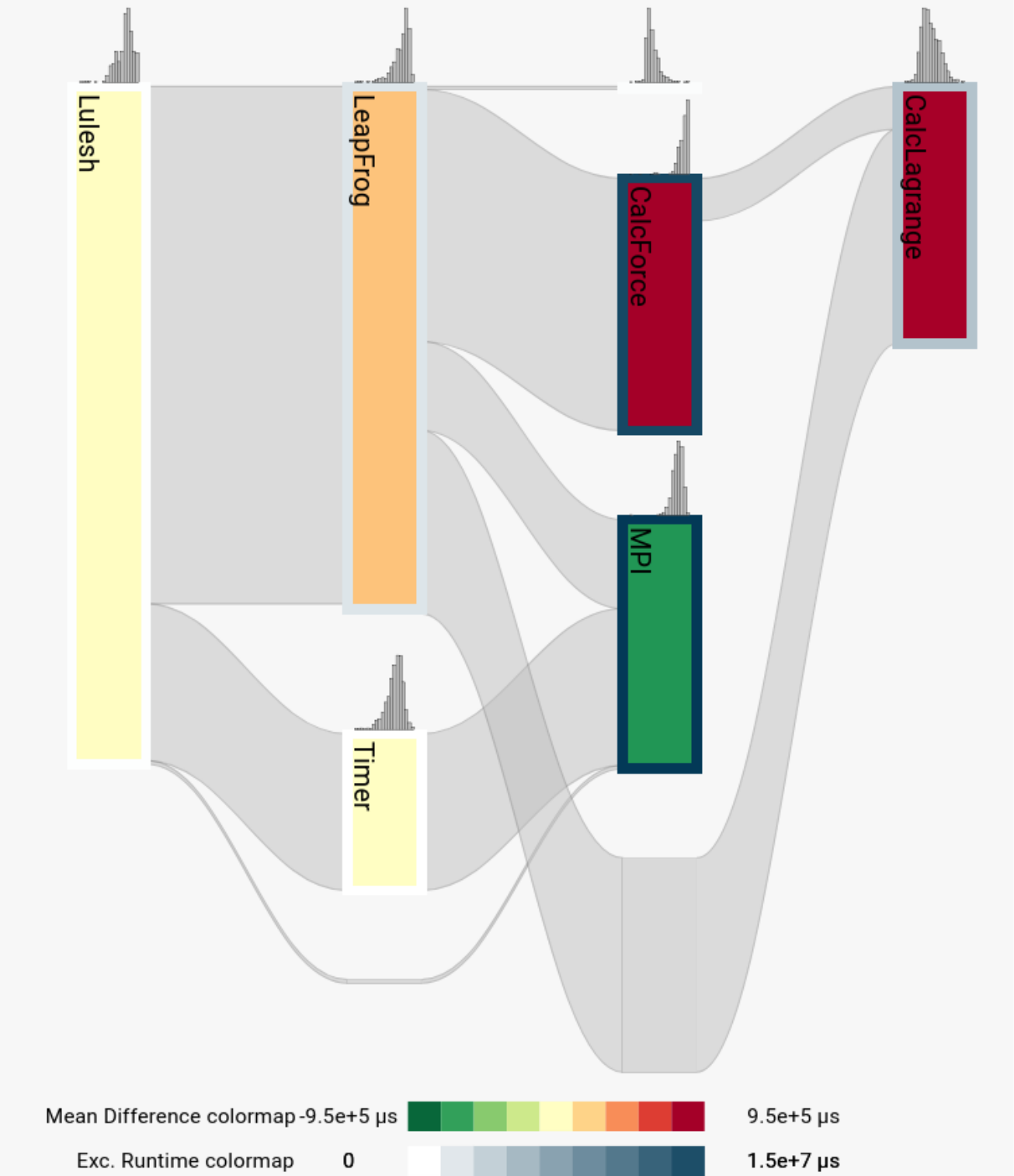}
        \label{fig:lulesh-diff1}
    }
    \hfill
    \subfloat[Diff view (216-core $-$ 125-cores)]{
        \includegraphics[height=5.8cm, keepaspectratio]{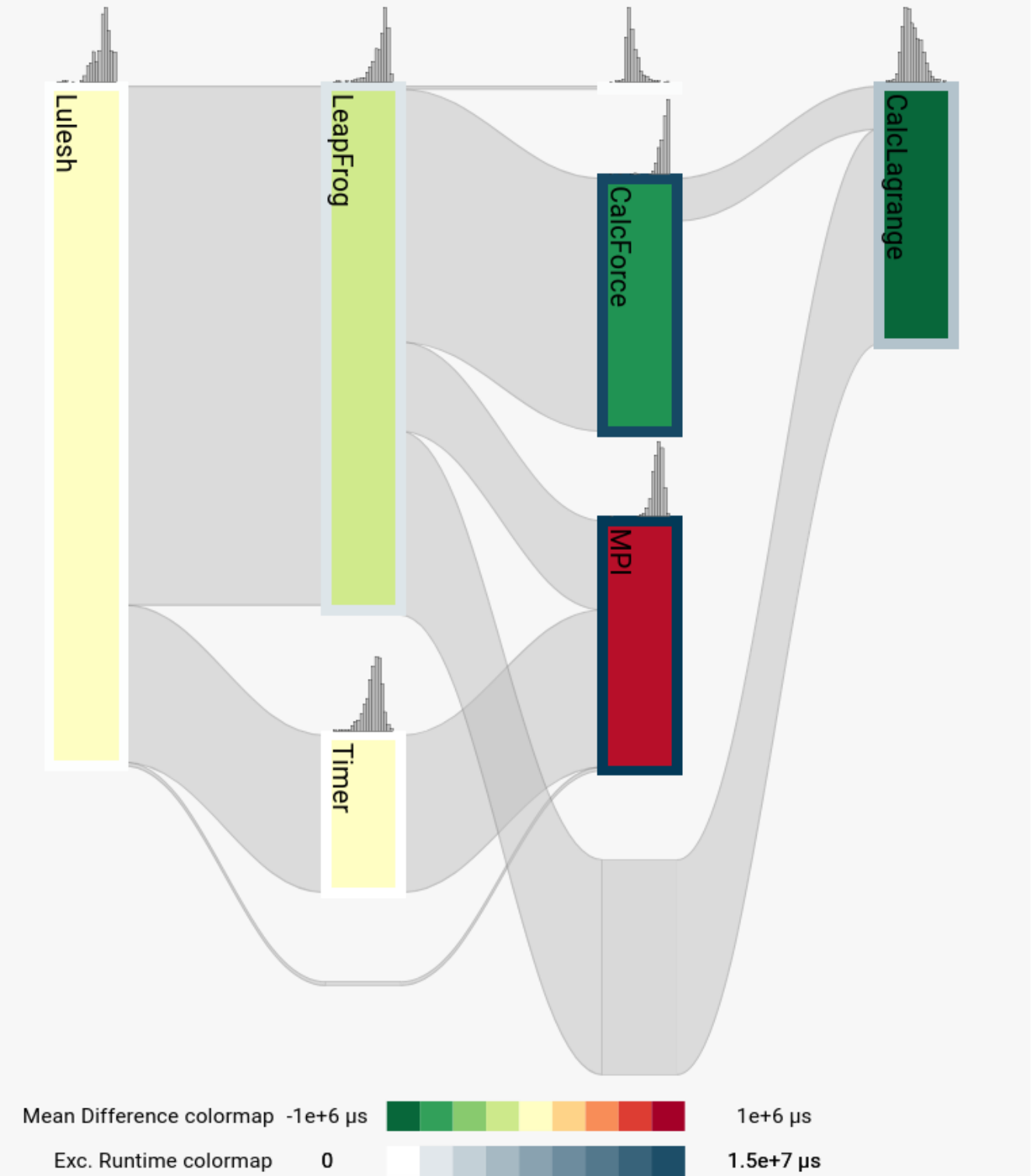}
        \label{fig:lulesh-diff2}
    }
    \hfill
  \vspace{-0.5em}    
\caption{Study of LULESH's performance profiles.
(a) The \textit{ensemble-gradients} and \textit{target-guides} 
        highlight the target run's runtime in contrast with the ensemble (the
        target run is 216-cores).
        (b) and (c) visualize pairwise differences of the runtimes between 2 runs 
        using a green-red colormap.
        Hues of red color highlight regions of code that cause a performance
        slowdown between the two profiles (\eg \pcode{CalcForce} and \pcode{CalcLagrange} in (b), and
        \pcode{MPI} in (c)), 
        and green hues highlight the performance speedup (\eg \pcode{MPI} in (b), and \pcode{CalcForce} and \pcode{CalcLagrange} in (c)).%
        \vspace{-0.75em}        
    } \label{fig:lulesh-compare}
\end{figure*}

%% file: f5_lulesh_summary.tex
\begin{figure*}[t]
\centering
%\fbox
{\includegraphics[width=0.85\textwidth]{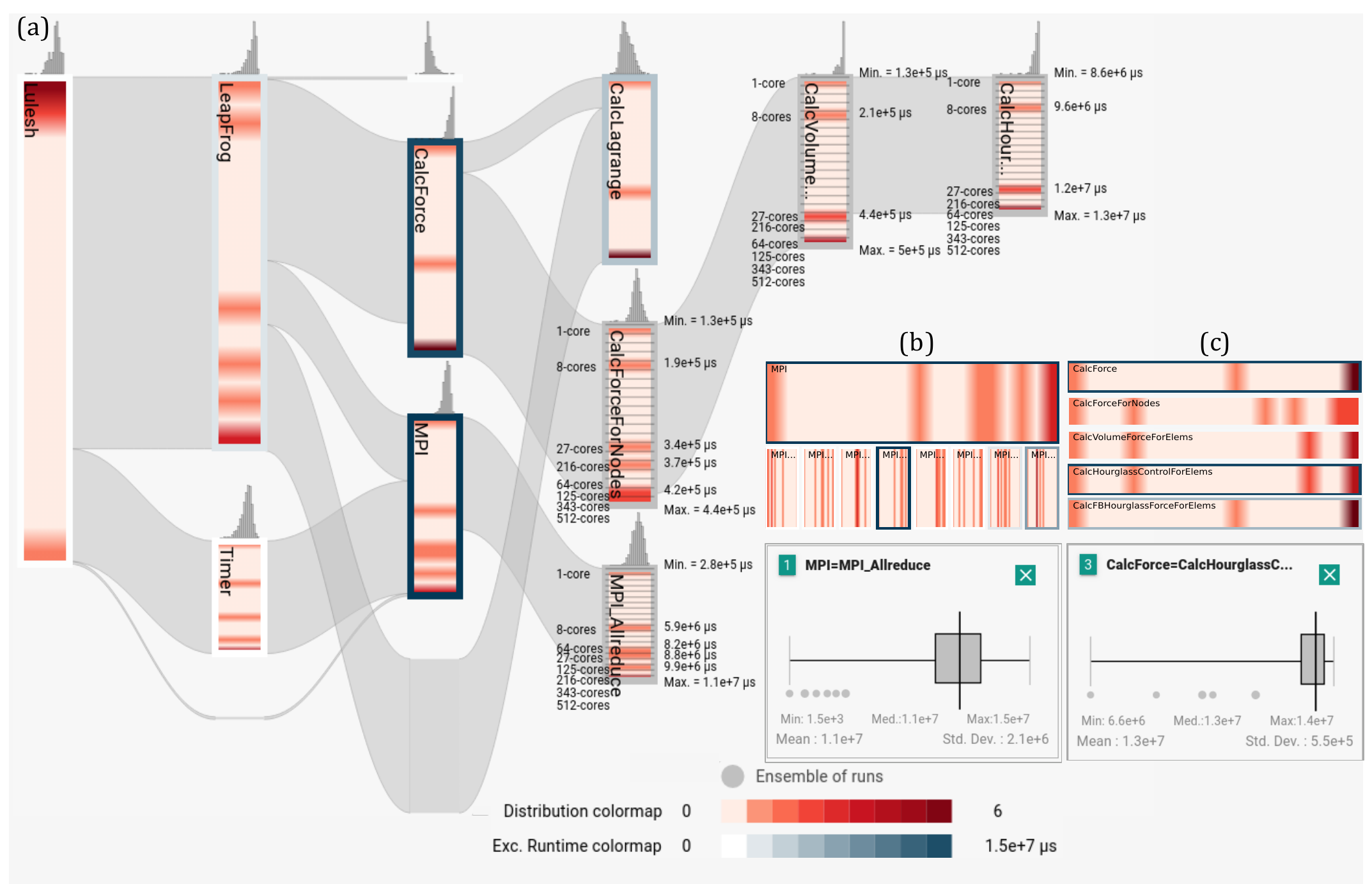}}
  \vspace{-1em}
    \caption{%
        (a) The exploration of LULESH profiles comprising of 8 runs using graph
        splitting operations and text labels reveal out-of-order runtimes (with respect
        to the scaling of resources). Further investigation through a sequence of
        splits as well as through (b) and (c) supernode hierarchy views lead to the
        problematic call sites: \pcode{CalcHourGlassControlforElems} and
        \pcode{MPI\_Allreduce}.
        %%
        % leads to the 
        %From \autoref{fig:lulesh-compare}(a), the calling contexts of the two expensive call sites, \pcode{MPI\_Allreduce} and \pcode{CalcHourGlassControlForElems} are revealed in the \gfse.
        %        %
        %        Encoding the exclusive runtime as border on the nodes inside the
        %        hierarchy of \pcode{MPI}(b) and \pcode{CalcForce}(c) reveals these two
        %        expensive call sites.
        %        %
        %        Below these hierarchies, the boxplots of the two call sites are shown to
        %        help understand the spread in their respective runtime distribution.
        \vspace{-1.5em}}\label{fig:lulesh-summary}
\end{figure*}

% The ensemble super graph
%     encodes the \textit{exclusive runtime} variability across 4 runs -- 27
%     cores, 64 cores, 125 cores, and 216 cores. The stroke color of each super
%     nodes encodes the runtime distribution. From this, we can identify the
%     expensive modules/libraries being \emph{CalcForce}(1.4e + 7 micro seconds)
%     and \emph{MPI} (1.5e + 7 micro seconds). Sorting the call site
%     correspondence view using the standard deviation of their mean exclusive
%     runtimes reveals that \emph{MPI Allreduce} and
%     \emph{CalcHourGlassControlForElems} are call sites with high variability.
%     These call sites are revealing using splitting interaction as shown in (a), The target guides are toggled for these nodes to understand their variation along their calling context. (b) and (c) show the calling structure as supernode
%     hierarchies for \emph{CalcForce} and \emph{MPI} module respectively.

%% file: 6_casestudy.tex
\section{Case Studies}
\label{casestudies}

To demonstrate and validate our visual analytics design, we present 3 case
studies illustrating its value to the HPC community.
% c) runtime and load balance issue between 100 runs,
%
%% ----------------------------------------------------------------------------
%% AMM case study
%% ----------------------------------------------------------------------------
\subsection{Performance Variability due to Application Parameters}

First, we study the performance of a single-process C++ library, AMM,
%~\cite{amm},
which creates adaptive representations on-the-fly for streaming volumetric
data.  AMM is currently under development and the code developers are keenly
interested in improving its performance for various application parameters.
Among others, a key parameter that affects the performance is the type of data
stream (\ie ordering of data, such as row-major and wavelet transform subband order)~\cite{Hoang:2019}.
With the general goal of obtaining insights into the performance variability of
AMM as well as identifying any potential optimization opportunities, experts
explore 18 profiles (captured through
Caliper~\cite{boehme2016caliper}) that represent other parametric variations
(\eg data sizes) for three different data streams.  \autoref{fig:amm} shows the
CallFlow visualization for the given ensemble.

Although AMM does not use MPI and therefore the corresponding profiles cannot
fully leverage the functionality of CallFlow, they make an excellent case study
due to high performance variability. Indeed, the ensemble view
(\autoref{fig:amm}) shows varied distributions of runtimes across different
run modes.
AMM makes use of several recursive functions, and despite different recursion
depths, the call graphs of the different runs are very similar. 
Given the application parameters explored here, code developers expected high
variation across runs but very similar patterns of variations across modules
and call sites.  Instead, to the surprise of code developers, the distribution
patterns (shown by ensemble gradients) vary significantly across different
modules.
This behavior hints at problems with recursive functions (\eg potential memory
leaks due to unnecessarily allocating new memory) -- a potential improvement in
the code that developers are currently exploring.

Through more detailed analysis of the ensembles, \eg using the supernode hierarchy
view and call site correspondence view, it was noted that several
\pcode{get\_*} functions in modules \pcode{AMTree} and \pcode{Octree} consume
significant run time. Through discussions with developers, it was learnt that
these functions make use of the default \pcode{std::unordered\_map}, and the
problem appears to be due to unnecessary hash collisions in the map. Developers
are currently experimenting with customizing the usage (\eg different hash and
different bucket counts) as well as a custom data structure to improve
performance.

Overall, AMM code developers were very impressed with CallFlow's capability to
easily and concisely describe the performance of the given ensemble and
identify potential bottlenecks. In general, performance profiling and
optimization is an ongoing process, and with the availability of our effective
visual analytic tool, developers will reevaluate the forthcoming development
versions of their code.

%% ----------------------------------------------------------------------------
%% LULESH case study
%% ----------------------------------------------------------------------------
\subsection{Performance Trends for a Weak Scaling Study}
\label{sec:lulesh-study}
\vspace{-0.1em}

% Identify slow callsites, for scaling bottlenecks 
% \textcolor{red}{Note: Comparing multiple executions that use a different number
%     of MPI processes. (Fig.4); A user might be interested in comparing two
%     executions that use a different number of MPI processes (Fig. 5 and Fig.
%     6);} \textcolor{red}{Compare a few call graphs. aggregate across processes
%     using means, and visualize the distribution. help understand the overall
%     behavior, show that it is easy to drill down into more details for nodes of
%     interest}

% % Lulesh - case study workflow:
% 1. Find performance variability among multi-rank executions.
% a. Load all 8 profiles. (1, 8, 27, 64, 125, 216, 343, and, 512 )
% b. Look at the PCA view. Have a quick examination of how the graphs are located in a higher-dimensional space.
% c. Select only multi-core executions using Lasso selection.
% d. Sort by standard deviation to see which callsite has maximum variability ( This can also be done by exploring the \dgE)

Most large-scale parallel applications are designed to utilize several
computational nodes and/or several cores per node.  Application developers and
HPC experts are often interested in scaling studies of such applications to
understand whether the codes are leveraging parallelism, \eg via MPI,
effectively.  Previously, a typical workflow for such use-cases would be the
user generating static charts, \eg bar plots to determine the changes in the
overall run time, perhaps at the module level or call site level. Usually, the
level of granularity is scripted in to generate the plot, and then analysis
performed, possible across levels of detail. Nevertheless, the lack of an
automated UI-based visualization imposes high time-to-insight, even for an
expert user.

Here, we are given an ensemble of multi-process performance profiles to study
weak scaling of a proxy application across eight execution parameters: 1, 8,
27, 64, 125, 216, 343, and 512 processes.  In particular, the application is
LULESH~\cite{karlin2012lulesh}, a Lagrangian shock hydrodynamics
mini-application that uses both MPI and OpenMP to achieve parallelism.
Recently, a similar case study was conducted by Bhatele
\etal~\cite{bhatele:sc2019} to study run-to-run performance differences for
identifying the most time-consuming regions of the code.  We use a similar
collection of profiles and showcase the advantages of visual analytics for such
exploration using CallFlow.

\para{Exploratory overview using ensemble-Sankey.}
The given profiles are converted into an ensemble super graph, which is shown
to the user using our ensemble-Sankey layout (see
\autoref{fig:lulesh-ensemble}).  Here, \gfse contains 33 call sites, which we
group into six modules.  Immediately, the user is conveyed the overall flow of
the resource (runtime) that establishes the context as the user can identify
the expected calling pattern.  The ensemble gradients on the different modules
of the visualization describe the \textit{complete} distribution, as compared
to a summary, \eg the mean value.  The interactivity of CallFlow allows the
user to select different ensemble members as target runs --- a functionality
particularly appreciated by the users because it allows comparing data in the
backdrop of a consistent context, \eg it is straightforward to compare the
``widths'' of the green (target) edges against the gray (ensemble) edges, in
\autoref{fig:lulesh-ensemble}.  Thus far, an initial exploration of the data
using the ensemble-view provides a good understanding of the collected profiles
--- not only qualitative but also quantitative (due to the associated labels and
colormaps).  Overall, although this is a small ensemble, the gradient patterns
largely indicate a reasonably good weak scaling.

\para{Pairwise comparison of profiles.}
In general, the users are often also interested in comparing pairs of profiles
within a larger ensemble, \eg a median profile vs.\ a slow profile.  This is
specifically true for the given data set as a curious behavior is observed: two
pairs of executions appear out of order in the ensemble-Sankey (labeled later
in \autoref{fig:lulesh-summary}).

Previously, Bhatele \etal~\cite{bhatele:sc2019} demonstrated their tool,
Hatchet, to analyze performance profiles and compute \emph{diff} between two
GraphFrames.
Using a color-mapped text-based tree visualization (similar to linux's
\emph{tree} command), Hatchet helps identify the nodes with positive/negative
values with respect to the diff.
% (see~\cite[Fig.~11]{bhatele:sc2019}).
%
Visualization of differences using colored text, however, imposes additional
perceptual complexity due to a lack of contrast among the colormapped values
with respect to the background.  Such visualization also suffers from
scalability issues and is static with no opportunity for interactive analysis.

Instead, the \emph{diff view} provided by CallFlow can reproduce such analysis
through a more-effective visual medium (see Figs.~\ref{fig:lulesh-diff1} and
\ref{fig:lulesh-diff2}).  The result highlights not only the modules that are
slower (with respect to the diff order) but also communicates the relative
degree of performance degradation easily.  For example, when scaling from 27 to
64 cores per node (\autoref{fig:lulesh-diff1}), \pcode{CalcForce} becomes about 
5\% more slower than \pcode{CalcLagrange}.  On the other hand, the
\pcode{MPI} module now takes longer when scaled from 125 to 216 cores
(\autoref{fig:lulesh-diff2}).  By encoding the pairwise differences onto the
ensemble graph, the visualization not only readily highlights the faster/slower
modules, but also allows the opportunity to interactively request more
information, \eg through additional views and/or graph splitting operations.
Furthermore, even though the \emph{diff view} is comparing only two graphs,
changing the pair through CallFlow UI allows the user to swiftly observe the
various pairwise differences within a consistent context (\ie the same ensemble
graph), irrespective of any fine-detail changes in the underlying CCT.

%% ------------------------------------------------------------------

\para{Comparing run-to-run slowdowns.}
Toward the ultimate goal of identifying call sites that exhibit inconsistent
and/or unexpected runtime behavior, further exploration is needed.  Here, by
toggling the text guides (using the click interaction), the runs corresponding
to the different bins in the ensemble gradient are listed along with the bin
value representing the gradient (see \autoref{fig:lulesh-compare}(a)).  The
text guides reveal two cases of out-of-order runtimes (with respect to
increasing core count) for \pcode{MPI} and \pcode{CalcForce} libraries.  One
way to highlight such differences is using the diff view
(\autoref{fig:lulesh-compare}).  Here, we further refine the ensemble view
using the interactive \emph{split graph} operation provided by CallFlow
(introduced in the previous version~\cite{nguyen2019callflow}) to reveal the
nodes inside these two modules.  Indeed, the text labels in the figure help
identify the culprit call site hierarchies,
[\pcode{MPI} $\to$ \pcode{MPI\_Allreduce}] and
    [\pcode{CalcForce} $\to$ \pcode{CalcForceForNodes}
        $\to$ \pcode{CalcVolumeForceForElems} $\to$ \pcode{CalcHourGlassControlForElems}].
A key advantage of using CallFlow is the immediate availability of alternate
views, \eg split ensemble view, supernode hierarchy view, and the call site
view (all shown in \autoref{fig:lulesh-compare}.  In particular, the supernode
hierarchy in \autoref{fig:lulesh-compare}(c) reasserts the upward propagation
of this behavior.  Finally, we look at the summarized distribution of the
respective call sites.  Even though the box plots highlight the outliers (with
respect to the interquartile range), by design, they are incapable of capturing
this anomalous behavior in the data (out-of-order runtimes) .  We take this
opportunity to emphasize that such behavior would also be missed even if the
user were to look at box-plot-type visualization outside CallFlow --- already a
stretch given the usual workflow.  Indeed, the true power of the visual
analytics enabled by CallFlow lies in its interactive linked views.

%% file: f6_osu_load.tex
% \begin{figure*}[!t]
%     \centering
%     %
%     \subfloat[24 runs of 12 MPI rank runs\label{fig:osu:24-runs}]
%     {\includegraphics[height=5cm]{Fig.7-osu-load/_google/Fig.6/Final/24-run-on-12-MPI-ranks}}
%     %
%     \hspace{1em}
%     \subfloat[b\label{fig:osu:24-runs}]
%     {\includegraphics[height=3cm]{Fig.7-osu-load/_google/Fig.6/Final/ensemble-histogram-view}}
%     %
%     %
%     \subfloat[Scaling to 100 runs\label{fig:osu:100runs}]
%     {\includegraphics[height=5cm, keepaspectratio]{Fig.7-osu-load/_google/Fig.6/Final/100-runs}}
%     %
%     \caption{}
%     \label{fig:osu}
% \end{figure*}

\begin{figure}[]
    \centering
    {\includegraphics[width=0.85\textwidth]{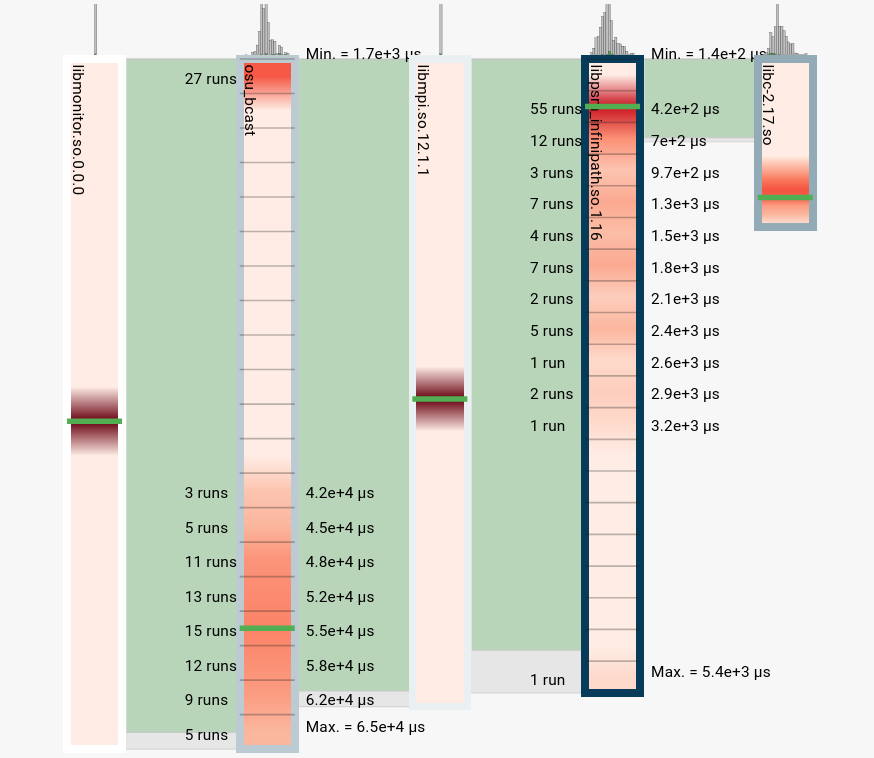}}
    \caption{CallFlow provides a scalable visual design that can accommodate 
    large ensembles. Here, a case study with 100 profiles is shown.
      \vspace{-2em} \label{fig:scalability}}
        \vspace{-0.8em}
\end{figure}

%% file: 7_discussion.tex
\section{Discussion}

\para{Conclusion.}
In this paper, we have extended CallFlow to create a scalable, interactive
visual analytic tool to study ensembles of call graphs.  Working closely with
domain experts, we identify the specific problems faced in the analysis of
collections of performance profiles, and map them to concrete comparative
visualization tasks using the framework of Gleicher
\etal~\cite{gleicher2017considerations}.  We develop a scalable visual
encoding, ensemble-Sankey, to describe the ensemble along with several other
linked visualizations. 

The proposed visual design itself is arbitrarily scalable in size of the
ensemble, since we show the distributions of runtimes for each supernode.
Nevertheless, practical concerns such as data processing and movement between
server and client poses limitations. We will improve the computational aspects
of the tool in the future versions.  Regardless, the tool itself can easily
scale up to about 100 profiles --- a moderately-sized use case, and still
maintain interactivity.  To showcase CallFlow's scalability to large
benchmarking experiments, we show (\autoref{fig:scalability}) an ensemble
of 100 profiles collected from an OSU\_Bcast benchmark across a range of MPI
processes (10, 12, 14, and 16).

\para{Expert feedback.}
Our collaborators comprise HPC researchers and application developers who also
participated in the discussion on the design of our system and provided
feedback.  In this work, we were able to successfully satisfy the requirements
arising from the domain problem, and have conducted a preliminary evaluation of
the visual tool with two case studies.  Overall the utility of the tool was
well received.  In general, getting an overall context of the ensemble is
integral to studying performance profiles.  It was noted that the new encoding,
the ensemble-Sankey, not only provides a new way of exploring the distributions
of interest, but is also remarkably straightforward and intuitive. Other linked
views, although not necessarily new visualization contributions, help tie in
the various pieces of information to put together a cohesive story about the
profiles that the user wishes to understand.

\para{Limitations and future work.}
Despite the positive reviews, our comparative visualization framework has
several directions for improvement.  For example, although the users expressed
excitement regarding the projection view in order to explore subsets of
ensembles, the lack of interpretability disallowed drawing any meaningful
insights. We will explore this idea further using tree-comparison
metrics instead.  More generally, with respect to visualization research, we currently
operated under the assumption of a ``reasonably'' similar ensemble. In the
future, we would like to explore visualizing graph ensembles without this
limitation.  Going forward, we also plan to explore the notion of ``ordering''
in ensembles, \eg performance profiles of software development, such as through
git history.  This additional complexity poses new challenges and it would be
interesting to see how a Sankey layout could be expanded to accommodate this
new dimension.